\documentclass[letterpaper,11pt,final]{article}

\usepackage[left=3cm,top=2cm,right=3cm,bottom=2cm]{geometry} 
\usepackage{multicol} 
\usepackage{url}
\usepackage{cuted}
\usepackage{flushend}

\usepackage[utf8]{inputenc} 
\usepackage[english]{babel} 
\usepackage{verbatim} 
\usepackage{color} 
\usepackage{chapterbib} 
\usepackage[normalem]{ulem}

\usepackage{amsmath}
\usepackage{amsfonts}
\usepackage{amssymb}
\usepackage{bbm}
\usepackage{cancel} 
\usepackage{youngtab}

\usepackage{graphicx}
\usepackage{feynmp} 
\usepackage{subfigure} 
\usepackage{float} 

\usepackage{multirow} 

\usepackage{pgfplots}
\usepackage{xparse}

\usetikzlibrary{positioning,arrows,patterns}
\usetikzlibrary{decorations.markings}
\usetikzlibrary{calc}

\tikzset{
  photon/.style={decorate, decoration={snake}, draw=black},
  fermion/.style={draw=black, postaction={decorate},decoration={markings,mark=at position .55 with {\arrow{>}}}},
  vertex/.style={draw,shape=circle,fill=black,minimum size=3pt,inner sep=0pt},
  vacuum/.style={draw,shape=rectangle,fill=black,minimum size=3pt,inner sep=0pt},
}

\NewDocumentCommand\semiloop{O{black}mmmO{}O{above}}
{%
\draw[#1] let \p1 = ($(#3)-(#2)$) in (#3) arc (#4:({#4+180}):({0.5*veclen(\x1,\y1)})node[midway, #6] {#5};)
}

\begin{document}

\title{$B$ meson decay anomaly with a non-universal $U(1)'$ extension}
\author{R. Martinez\thanks{remartinezm@unal.edu.co}, F. Ochoa\thanks{faochoap@unal.edu.co}, J.M. Quimbayo\thanks{jmquimbayog@unal.edu.co},  }

\maketitle
\begin{center}
\textit{Departamento de F\'{i}sica, Universidad Nacional de Colombia,
Ciudad Universitaria, K. 45 No. 26-85, Bogot\'a D.C., Colombia} 
\par\end{center}



\vspace*{0cm}
\begin{center}\rule{0.9\textwidth}{0.1mm} \end{center}
\begin{abstract}
We propose an extension of the standard model with an extra $U(1)^{\prime}$ abelian symmetry, three Higgs doublets and two Higgs singlets, where the new $U(1)'$ charges are flavour non-universal. As a result, the model introduces an enlarger particle spectrum in the TeV scale with large new physics possibilities. The model reproduces the mixing angles and mass structures of the quarks, charged and neutral leptons. We found scenarios where the observed anomaly of the $B^{+}\rightarrow K^+ \ell \ell$ decay can be explained due to the existence of couplings with new extra fermions at the TeV scale. By modifying the parametrization of the mixing matrices, we found solution in the decoupling limit.
 
\begin{center}\rule{0.9\textwidth}{0.1mm} \end{center}
\vspace*{0.5cm}
\end{abstract}

\section{Introduction}
Despite all its success, the Standard Model (SM) of Glashow, Weinberg and Salam \cite{SM} does not account for all the theoretical and experimental observations; therefore it is believed that there is a more fundamental theory where the SM emerges as an effective lower limit at the electroweak scale. For example, the fermion mass hierarchy and the neutrino mass problem are two related subjects that may be understood as manifestations of an underlying theory beyond the SM \cite{hierarchy, neutrino}. Also, there are some observables that show some tensions from the SM predictions, which may be associated to new physics. Among the different observations accesible to collider physics, the flavor observables imposes stringent limits to many SM extensions. In particular, the lepton universality exhibited by the SM is sensitive to new physics that can be tested in rare semi-leptonic transitions of mesons \cite{semileptonic, semi-2, anomaly}. Recently, the ratio of the branching fractions of B meson decays into muon and electron pairs was reported by the LHCb collaboration \cite{lhcb}, where a deviation from the SM prediction within $2.6 \sigma$ suggest us a possible lepton universality violation not explained in the framework of the SM.

From the theoretical point of view, the mass hierarchy problem can be addressed in a model independent approach by assuming texture zero structures for the mass matrices \cite{zerotexture}. Relations between mixing angles and masses can be also derived for both quarks and lepton sectors in models with broken flavor symmetries. These symmetries, that relates the three fermion families in a non-trivial way, have been extensively studied in the literature in different extensions of the SM, either as a continuos abelian or non-abelian extensions  \cite{continuous}, or as a discrete flavour symmetry \cite{discrete}. On the other hand, many grand unified and superstring models predicts one or multiple extra abelian symmetries in their effective low energy limit \cite{unified}, which motivates extensions of the SM with an extra U(1)' gauge symmetry. If this symmetry is family nonuniversal, it is possible to connect the flavour problem with the group properties of these models. Also, these type of extensions imply a new extra neutral $Z'$ boson which, in the framework of a nonuniversal model, produces new contributions to flavour-changing neutral current (FCNC) processes. Although these type of interactions are strongly suppressed in the SM, which is consistent with most of the experimental observations, there are some anomalies reported, as the aforementioned decay $B^+\rightarrow K^+\ell^+\ell^-$, corresponding to a $b\rightarrow s$ FCNC process.    

Motivated initially by the mass hierarchy of the quarks, in the U(1)' model proposed in reference \cite{extra-u1}, the new abelian charge distinguishes one family of quarks from the other two, and newly vector-like quarks are introduced in order to restore the cancellation of the chiral anomalies. This model is universal in the lepton sector. A modification of the above model was presented in \cite{lepton-nonuniversal}, where newly charged leptons are also introduced in order to obtain nonuniversal lepton families compatible with the cancellation of the chiral anomalies. The above structure can reproduce the elements of the mixing mass matrix and the squared-mass diferences data from neutrinos oscillations experiments. 

In this article, we combine the nonuniversal $U(1)'$ model from \cite{lepton-nonuniversal} with a three higgs doublet model in order to explain the anomaly measured by the LHCb in the ratio of the branching fractions of $B^+\rightarrow K^+\mu^+\mu^-$ to $B^+\rightarrow K^+e^+e^-$ decays. This model was already presented in \cite{1s3d-nonuniversal}, where the three doublets may induce naturally the fermion hierarchy, while the smallness of the neutrinos can be implemented through an inverse see saw mechanism, where the vacuum expectation value that breaks the extra $U(1)'$ symmetry defines the large mass scale.

\section{Survey of the model}

\subsection{Particle content}

\begin{table*}
\centering
\begin{tabular}{cc|cc|cc}
\hline\hline
Bosons	&	$X^{\pm}$	&	Quarks	&	$X^{\pm}$	&	Leptons	&	$X^{\pm}$	\\ \hline 
\multicolumn{2}{c}{Scalar Doublets}	&
\multicolumn{4}{c}{SM Fermionic Doublets}	\\ \hline\hline
$\Phi_{1}=\left(\begin{array}{c}
\phi_{1}^{+} \\ \frac{h_{1}+v_{1}+i\eta_{1}}{\sqrt{2}}	
\end{array}\right)$	&	$\frac{+2}{3}^{+}$	&
$q^{1}_{L}=\left(\begin{array}{c}u^{1}	\\ d^{1} \end{array}\right)_{L}$
	&	$\frac{+1}{3}^{+}$	&	
$\ell^{e}_{L}=\left(\begin{array}{c}\nu^{e} \\ e^{e} \end{array}\right)_{L}$
	&	$0^{+}$	\\
$\Phi_{2}=\left(\begin{array}{c}
\phi_{2}^{+} \\ \frac{h_{2}+v_{2}+i\eta_{2}}{\sqrt{2}}	
\end{array}\right)$	&	$\frac{+1}{3}^{-}$	&
$q^{2}_{L}=\left(\begin{array}{c}u^{2} \\ d^{2} \end{array}\right)_{L}$
	&	$0^{-}$	&
$\ell^{\mu}_{L}=\left(\begin{array}{c}\nu^{\mu} \\ e^{\mu} \end{array}\right)_{L}$
	&	$0^{+}$		\\
$\Phi_{3}=\left(\begin{array}{c}
\phi_{3}^{+} \\ \frac{h_{3}+v_{3}+i\eta_{3}}{\sqrt{2}}	
\end{array}\right)$	&	$\frac{+1}{3}^{+}$	&
$q^{3}_{L}=\left(\begin{array}{c}u^{3} \\ d^{3} \end{array}\right)_{L}$
	&	$0^{+}$	&
$\ell^{\tau}_{L}=\left(\begin{array}{c}\nu^{\tau} \\ e^{\tau} \end{array}\right)_{L}$
	&	$-1^{+}$	\\   \hline\hline
	
\multicolumn{2}{c}{Scalar Singlets} &\multicolumn{4}{c}{SM Fermionic Singlets}	\\ \hline\hline
\begin{tabular}{c}$\chi  =\frac{\xi_{\chi}  +v_{\chi}  +i\zeta_{\chi}}{\sqrt{2}}$\\$\sigma$\end{tabular}	&
\begin{tabular}{c}$\frac{-1}{3}^{+}$\\$\frac{-1}{3}^{-}$\end{tabular}	&
\begin{tabular}{c}$u_{R}^{1,3}$\\$u_{R}^{2}$\\$d_{R}^{1,2,3}$\end{tabular}	&	 
\begin{tabular}{c}$\frac{+2}{3}^{+}$\\$\frac{+2}{3}^{-}$\\$\frac{-1}{3}^{-}$\end{tabular}	&
\begin{tabular}{c}$e_{R}^{e}$\\$e_{R}^{\mu}$\\$e_{R}^{\tau}$\end{tabular}	&	
\begin{tabular}{c}$\frac{-4}{3}^{+}$\\$\frac{-1}{3}^{+}$\\$\frac{-4}{3}^{-}$\end{tabular}\\   \hline \hline 

\multicolumn{2}{c}{Gauge bosons}&	\multicolumn{2}{c}{Non-SM Quarks}&	\multicolumn{2}{c}{Non-SM Leptons}\\ \hline \hline
\begin{tabular}{c}$W^{\pm}_{\mu}$\\$W^{3}_{\mu}$\end{tabular}	&
\begin{tabular}{c}$0^{+}$\\$0^{+}$\end{tabular}	&
\begin{tabular}{c}$\mathcal{T}_{L}$\\$\mathcal{T}_{R}$\end{tabular}	&
\begin{tabular}{c}$\frac{+1}{3}^{-}$\\$\frac{+2}{3}^{-}$\end{tabular}	&
\begin{tabular}{c}$\nu_{R}^{1,2,3}$\\$\mathcal{N}_{R}^{1,2,3}$\end{tabular} 	&	
\begin{tabular}{c}$\frac{+1}{3}^{+}$\\$0^{+}$\end{tabular}\\
$B_{\mu}$	&	  $0^{+}$	&	
$\mathcal{J}^{1,2}_{L}$	&	  $0^{+}$	&
$\mathcal{E}_{L}^{1},\mathcal{E}_{R}^{2}$	&	$-1^{+}$	\\

$Z'_{\mu}$	&	 $0^{+}$	&
$\mathcal{J}^{1,2}_{R}$	&	 $\frac{-1}{3}^{+}$	&
$\mathcal{E}_{R}^{1},\mathcal{E}_{L}^{2}$	&	$\frac{-2}{3}^{+}$	\\ \hline \hline
\end{tabular}
\caption{Non-universal $X$ quantum number and $\mathbb{Z}_{2}$ parity for SM and non-SM fermions.}
\label{tab:Particle-content}
\end{table*}

The model is an extension of the SM group, with the addition of a nonuniversal abelian gauge symmetry (which we denote as $\mathrm{\mathrm{U}(1)}_{X}$) whose gauge boson and coupling constant are $Z_{\mu}'$ and $g_{X}$, respectively, while the weak hipercharge is defined as usual through the Gell-Mann-Nishijima relation:

\begin{eqnarray}
Q=I_3+\frac{Y}{2},
\end{eqnarray}
with $Q$ the electric charge operator and $I_3$ the isospin.

The additional gauge symmetry introduces new chiral anomaly equations which can be solved by assigning non-trivial $X$-quantum numbers to the fermions of the SM \cite{extra-u1}. If this new $U(1)$ charges are different from the SM $U(1)_Y$ charges, then anomaly cancellation require new quarks and leptons to be added into the spectrum with masses at a larger scale than the electroweak scale \cite{dobrescu}. For simplicity, all the new particles are assumed to be singlets under the gauge $SU(2)_L$ group. In order to provide masses to the new sector, we introduce a neutral Higgs singlet $\chi$ with non-vanishing VEV, and $\mathrm{\mathrm{U}(1)}_{X}$ charge $X=-1/3$ 
in such a way that it spontaneously breaks the new gauge symmetry. Another scalar singlet $\sigma$ identical to $\chi$ but without VEV is introduced, which can play the role of a dark matter candidate, such as in the $U(1)'$ extension in \cite{dilepton}. 

On the other hand, the phenomenological fermions define three mass scales. First, the top quark ($t$) is the heaviest observed femion at $10^2$ GeV scale. Second, the tau lepton and bottom quark $(\tau ,b)$ are in the $10^0$ GeV scale. Finally, the muon and strange quark $(\mu, s)$ are in the $10^2$ MeV scale. This general structure can be directly induced with three Higgs doublets with vacuum expectation values (VEV) $v_1 >  v_2  > v_3$ associated to the three scales above. The chosen particle spectrum is presented in the table \ref{tab:Particle-content} where three new quarks ($\mathcal{T}$, $\mathcal{J}^{1,2}$) and two charged leptons ($\mathcal{E}^{1,2}$) are introduced. To obtain masses for the active neutrinos, we introduce three right-handed neutrinos $\nu_{R}^{1,2,3}$ with nontirivial $U(1)_X$ charges, which allow the coupling with the ordinary lepton doublets $\ell _L$, but because of their $X-$charge, they do not generate Majorana mass terms. The addition of Majorana fermions, $\mathcal{N}_R^{1,2,3}$, allow an inverse see saw mechanism in order to explain the smallness of the active neutrinos. 

In order to obtain predictable and analitycal relations for the masses and mixing angles of the fermions according to observations, we assign specific $\mathbb{Z}_{2}$ symmetry parities, which are shown as superscripts in the $X$-charges. It is to note that, despite the scalar doublets $\Phi_{2}$ and $\Phi_{3}$ have the same $X$ charge, they have opposite $\mathbb{Z}_{2}$ parity such that their couplings to fermions are complementary. 

Finally, since the new fermions are vector-like under the SM group, they do not introduce any extra $SU(2)_L\times U(1)_Y$ contribution to the anomaly equations. However, the symmetry $U(1)_{X}$ may generate the following pure and mixed anomalies:

\begin{eqnarray}
\left[ U(1)_{X}\right] ^{3} &\rightarrow &A_{1}=
\sum _{\ell ,Q} \left[X_{\ell _{L}}^3+3X_{Q_{L}}^3\right]-\sum _{\ell ,Q} \left[X_{\ell _{R}}^3+3X_{Q_{R}}^3\right] \notag
\\
\left[ SU(3)_{c}\right] ^{2} U(1)_{X} &\rightarrow &A_{2}=
\sum _{Q} X_{Q_{L}}-\sum_{Q}X_{Q_R}  \notag 
\\
\left[ SU(2)_{L}\right] ^{2} U(1)_{X} &\rightarrow&A_{3}=
\sum _{\ell} X_{\ell _{L}}+3\sum_{Q}X_{Q_L},  \notag 
\\
\left[ U(1)_{Y}\right] ^{2} U(1)_{X} &\rightarrow&A_{4}=
\sum _{\ell ,Q} \left[Y_{\ell _L}^2X_{\ell _{L}}+3Y_{Q_L}^2X_{Q_{L}}\right]-\sum _{\ell ,Q} \left[Y_{\ell _R}^2X_{\ell _{R}}+3Y_{Q_R}^2X_{Q_{R}}\right]  \notag 
\\
U(1)_{Y} \left[ U(1)_{X} \right] ^{2}&\rightarrow&A_{5}=
\sum _{\ell ,Q} \left[Y_{\ell _L}X_{\ell _{L}}^2+3Y_{Q_L}X_{Q_{L}}^2\right]-\sum _{\ell ,Q} \left[Y_{\ell _R}X_{\ell _{R}}^2+3Y_{Q_R}X_{Q_{R}}^2\right]  \notag 
\\
\left[ Grav\right] ^{2}\otimes U(1)_{X} &\rightarrow&A_{6}=
\sum _{\ell ,Q} \left[X_{\ell _{L}}+3X_{Q_{L}}\right]-\sum _{\ell ,Q} \left[X_{\ell _{R}}+3X_{Q_{R}}\right] 
\label{anomalias}
\end{eqnarray}
where the sums in $Q$ runs over all the quarks, and $\ell$ over all leptons. However, by direct calculation, it is possible to verify that the chosen $U(1)_X$ charges satisfy the cancellation of the anomalies in (\ref{anomalias}), so that the model is free from chiral anomalies. 


\subsection{Lagrangians}

\subsubsection*{Yukawa interactions}

The most general Yukawa Lagrangian must obey the gauge symmetry $G_{SM}\times U(1)_X$ in order to obtain a renomarlizable model, where $G_{SM}$ is the SM gauge group. However, we impose additionaly that the interactions respect the discrete $Z_2$ symmetry, where each particle has the intrisic $Z_2$-parity shown in table \ref{tab:Particle-content}. Since there are particles with different $Z_2$-parities, not all couplings between fermions and scalars are allowed. Specifically, the Yukawa Lagrangian allowed by the symmetries of the model for the up- and down-like quarks are:

\begin{equation}
\begin{split}
-\mathcal{L}_{U} &= 
h_{3 u}^{1 1}\overline{q_{L}^{1}}\tilde{\Phi}_{3}u_{R}^{1} + 
h_{2 u}^{1 2}\overline{q_{L}^{1}}\tilde{\Phi}_{2}u_{R}^{2} + 
h_{3 u}^{1 3}\overline{q_{L}^{1}}\tilde{\Phi}_{3}u_{R}^{3} + 
h_{1 u}^{2 2}\overline{q_{L}^{2}}\tilde{\Phi}_{1}u_{R}^{2} \\ &+ 
h_{1 u}^{3 1}\overline{q_{L}^{3}}\tilde{\Phi}_{1}u_{R}^{1} + 
h_{1 u}^{3 3}\overline{q_{L}^{3}}\tilde{\Phi}_{1}u_{R}^{3} +
h_{2 \mathcal{T}}^{1} \overline{q_{L}^{1}}\tilde{\Phi}_{2}\mathcal{T}_{R} +
h_{1 \mathcal{T}}^{2} \overline{q_{L}^{2}}\tilde{\Phi}_{1}\mathcal{T}_{R} \\ &+
g_{\sigma u}^{1}\overline{\mathcal{T}_{L}}\sigma u_{R}^{1} + 
g_{\chi u}^{2}\overline{\mathcal{T}_{L}}\chi u_{R}^{2} \,+ 
g_{\sigma u}^{3}\overline{\mathcal{T}_{L}}\sigma u_{R}^{3} \,+ 
g_{\chi \mathcal{T}}\overline{\mathcal{T}_{L}}\chi \mathcal{T}_{R} + \mathrm{h.c.},
\end{split}
\label{eq:Up-Lagrangian}
\end{equation}
\begin{equation}
\begin{split}
-\mathcal{L}_{D} &= 
h_{1 \mathcal{J}}^{1 1}\overline{q_{L}^{1}}{\Phi}_{1}\mathcal{J}_{R}^{1} + 
h_{2 \mathcal{J}}^{2 1}\overline{q_{L}^{2}}{\Phi}_{2}\mathcal{J}_{R}^{1} + 
h_{3 \mathcal{J}}^{3 1}\overline{q_{L}^{3}}{\Phi}_{3}\mathcal{J}_{R}^{1} +
h_{1 \mathcal{J}}^{1 2}\overline{q_{L}^{1}}{\Phi}_{1}\mathcal{J}_{R}^{2} \\ &+ 
h_{2 \mathcal{J}}^{2 2}\overline{q_{L}^{2}}{\Phi}_{2}\mathcal{J}_{R}^{2} + 
h_{3 \mathcal{J}}^{3 2}\overline{q_{L}^{3}}{\Phi}_{3}\mathcal{J}_{R}^{2} + 
h_{3 d}^{2 1}\overline{q_{L}^{2}}{\Phi}_{3}d_{R}^{1} +
h_{3 d}^{2 2}\overline{q_{L}^{2}}{\Phi}_{3}d_{R}^{2} \\ &+
h_{3 d}^{2 3}\overline{q_{L}^{2}}{\Phi}_{3}d_{R}^{3} +
h_{2 d}^{3 1}\overline{q_{L}^{3}}{\Phi}_{2}d_{R}^{1} +
h_{2 d}^{3 2}\overline{q_{L}^{3}}{\Phi}_{2}d_{R}^{2} +
h_{2 d}^{3 3}\overline{q_{L}^{3}}{\Phi}_{2}d_{R}^{3} \\ &+ 
g_{\sigma d}^{1 1}\overline{\mathcal{J}_{L}^{1}}\sigma^{*} d_{R}^{1} + 
g_{\sigma d}^{1 1}\overline{\mathcal{J}_{L}^{1}}\sigma^{*} d_{R}^{2} +
g_{\sigma d}^{1 3}\overline{\mathcal{J}_{L}^{1}}\sigma^{*} d_{R}^{3} + 
g_{\sigma d}^{2 1}\overline{\mathcal{J}_{L}^{2}}\sigma^{*} d_{R}^{1} \\ &+
g_{\sigma d}^{2 2}\overline{\mathcal{J}_{L}^{2}}\sigma^{*} d_{R}^{2} +
g_{\sigma d}^{2 3}\overline{\mathcal{J}_{L}^{2}}\sigma^{*} d_{R}^{3} + 
g_{\chi \mathcal{J}}^{1}\overline{\mathcal{J}_{L}^{1}}\chi^{*} \mathcal{J}_{R}^{1} + 
g_{\chi \mathcal{J}}^{2}\overline{\mathcal{J}_{L}^{2}}\chi^{*} \mathcal{J}_{R}^{2} + \mathrm{h.c.},
\end{split}
\label{eq:Down-Lagrangian}
\end{equation}
while for the neutral and charged leptons we obtain:

\begin{equation}
\begin{split}
-\mathcal{L}_{N} &=
h_{3 \nu}^{e e}\overline{\ell^{e}_{L}}\tilde{\Phi}_{3}\nu^{1}_{R} + 
h_{3 \nu}^{e \mu}\overline{\ell^{e}_{L}}\tilde{\Phi}_{3}\nu^{2}_{R} + 
h_{3 \nu}^{e \tau}\overline{\ell^{e}_{L}}\tilde{\Phi}_{3}\nu^{3}_{R} +
h_{3 \nu}^{\mu e}\overline{\ell^{\mu}_{L}}\tilde{\Phi}_{3}\nu^{1}_{R} \\ &+
h_{3 \nu}^{\mu \mu}\overline{\ell^{\mu}_{L}}\tilde{\Phi}_{3}\nu^{2}_{R} + 
h_{3 \nu}^{\mu \tau}\overline{\ell^{\mu}_{L}}\tilde{\Phi}_{3}\nu^{3}_{R} +
g_{\chi \mathcal{N}}^{i j} \overline{\nu_{R}^{i\;C}} \chi^{*} \mathcal{N}_{R}^{j} +
\frac{1}{2} \overline{\mathcal{N}_{R}^{i\;C}} M^{ij}_{\mathcal{N}} \mathcal{N}_{R}^{j} + \mathrm{h.c.},
\end{split}
\label{eq:Neutrino-Lagrangian}
\end{equation}
\begin{equation}
\begin{split}
-\mathcal{L}_{E} &=
h_{3 e}^{e \mu}\overline{\ell^{e}_{L}}\Phi_{3}e^{\mu}_{R} + 
h_{3 e}^{\mu \mu}\overline{\ell^{\mu}_{L}}\Phi_{3}e^{\mu}_{R} + 
h_{2 e}^{\tau e}\overline{\ell^{\tau}_{L}}\Phi_{3}e^{e}_{R} +
h_{2 e}^{\tau \tau}\overline{\ell^{\tau}_{L}}\Phi_{2}e^{\tau}_{R} \\ &+
h_{1 E}^{e 1}\overline{\ell^{e}_{L}}\Phi_{1}\mathcal{E}_{R}^{1} + 
h_{1 \mathcal{E}}^{\mu 1}\overline{\ell^{\mu}_{L}}\Phi_{1}\mathcal{E}_{R}^{1} +
g_{\chi e}^{1 e}\overline{\mathcal{E}_{L}^{1}}\chi^{*} e^{e}_{R} + 
g_{\chi e}^{2 \mu}\overline{\mathcal{E}_{L}^{2}}\chi e^{\mu}_{R} + \\ &+
g_{\chi \mathcal{E}}^{1}\overline{\mathcal{E}_{L}^{1}}\chi \mathcal{E}_{R}^{1} + 
g_{\chi \mathcal{E}}^{2}\overline{\mathcal{E}_{L}^{2}}\chi^{*} \mathcal{E}_{R}^{2} + \mathrm{h.c.},
\end{split}
\label{eq:Electron-Lagrangian}
\end{equation}
where $\widetilde{\Phi}=i\sigma_2 \Phi^*$ are the scalar doublet conjugates and the Majorana mass components are denoted as $M^{ij}_{\mathcal{N}}$.

\subsubsection*{Gauge and scalar boson interactions}

The Higgs kinetic Lagrangian contains the couplings among vector gauge and scalar bosons, which takes the general form

\begin{eqnarray}
\mathcal{L}_{kin} &=& \left(D_{\mu }S\right)^{\dag}\left(D^{\mu }S \right),
\label{eq:Kinetic-Lagrangian}
\end{eqnarray}
where the covariant derivative is defined as:

\begin{eqnarray}
D^{\mu }=\partial ^{\mu}-igW^{\mu}_{\alpha}T_S^{\alpha}-ig'\frac{Y_S}{2}B^{\mu}-ig_XX_S Z'^{\mu}.
\label{Covariant}
\end{eqnarray}
The parameters $2T_S^{\alpha}$ corresponds to the Pauli matrices when $S=\Phi _{1,2,3}$ and $T_S^{\alpha}=0$ when $S=\chi , \sigma$, while $Y_S$ and $X_S$ correspond to the hypercharge and $U(1)_X$ charge according to the values in table \ref{tab:Particle-content}. The gauge coupling constants $g$ and $g'$ obey the same relation as in the SM, $g'=g\tan \theta _W$, with $\theta _W$ the Weinberg angle.


\subsubsection*{Dirac Lagrangian}

Finally, the interactions of fermions through vector gauge fields are described by the following Lagrangian:

\begin{eqnarray}
\mathcal{L}_{D} &=&\text{i}\overline{f_{Li}}\gamma ^{\mu}D_{\mu}f_{Li}+\text{i}\overline{f_{Ri}}\gamma ^{\mu}D_{\mu}f_{Ri},
\label{Dirac-Lagrangian}
\end{eqnarray}
where $f_i$ runs over all flavour of fermions, and, as usual, a sum over repeated indices is implied. The covariant derivative $D^{\mu}$ is similar to (\ref{Covariant}) but changing the scalar parameters by the corresponding fermion parameters.

\subsection{Mass eigenstates and interactions}

\subsubsection*{Fermion masses}

The Yukawa Lagrangians from (\ref{eq:Up-Lagrangian}) to (\ref{eq:Electron-Lagrangian}) provide masses to all the fermions after the symmetries of the model breaks spontaneously, through the vacuum structure of the Higgs fields shown in table \ref{tab:Particle-content}. In general, the mass terms have the following form:

\begin{eqnarray}
-\mathcal{L}_{f} &=&\overline{\mathbf{f}_L}M_f\mathbf{f}_R+\text{h.c.},
\label{mass-lagrangian}
\end{eqnarray}
where $\mathbf{f}$ are fermion multiplets with components of the same electric charge, namely     

\begin{eqnarray}
\mathbf{f}&:& \mathbf{U}=(u^1,u^2,u^3,\mathcal{T}) \nonumber \\
&& \mathbf{D}=(d^1,d^2,d^3,\mathcal{J}^1, \mathcal{J}^2 ) \nonumber \\
&& \mathbf{E}=(e^e,e^{\mu},e^{\tau},\mathcal{E}^1, \mathcal{E}^2 ) \nonumber \\
&& \mathbf{N}_L=(\nu_L^{e, \mu, \tau},\nu_R^{1,2,3C},\mathcal{N}_R^{1,2,3C} ),
\label{flavour-multilets}
\end{eqnarray}
and $M_f$ are complex non-diagonal mass matrices. In general, the above mass matrices can be diagonalized by bi-unitary transformations of the form:

\begin{eqnarray}
m_f=\left(V_L^{f}\right)^{\dag}M_fV_R^{f},
\label{fermion-mass-diagonal}
\end{eqnarray}
which, after replacing in (\ref{mass-lagrangian}), lead us to the left- and right-handed mass basis:

\begin{eqnarray}
\mathbf{\tilde f}_L=\left(V_L^{f}\right)^{\dag}\mathbf{f}_L, \ \ \ \ \ \mathbf{\tilde f}_R=\left(V_R^{f}\right)^{\dag}\mathbf{f}_R,
\label{fermion-mass-basis}
\end{eqnarray}
where:

\begin{eqnarray}
\mathbf{\tilde f}&:& \mathbf{\tilde U}=(u, c, t, T) \nonumber \\
&& \mathbf{\tilde D}=(d, s, b, J^1, J^2 ) \nonumber \\
&& \mathbf{\tilde E}=(e, \mu, \tau, E^1, E^2 ) \nonumber \\
&& \mathbf{\tilde N}_L=(\nu_L^{1, 2, 3}, \tilde \nu_R^{1,2,3C}, N_R^{1,2,3C} ),
\label{mass-multilets}
\end{eqnarray}

The specific form of the matrices $V_{L,R}^{f}$ depends on the Yukawa structure of the original Lagrangians in  (\ref{eq:Up-Lagrangian})-(\ref{eq:Electron-Lagrangian}). In particular, with the choosen $Z_2$-parities, these Yukawa terms lead us to predictible mass structures for quarks, charged leptons and neutrinos, as shown in \cite{1s3d-nonuniversal}, which we summarize in the appendix \ref{app:biunitary}.  

\subsubsection*{The unitary constraint}

Each rotation matrix in (\ref{fermion-mass-basis}) must obey the unitary condition

\begin{eqnarray}
\left(V_{L,R}^{f}\right)^{\dag}V_{L,R}^{f}=I,
\label{unitarity}
\end{eqnarray}
where $I$ is the identity. In the above relation, we must take into account that the sum from the matrix products contain two contributions due to the components with ordinary SM particles and the newly vector-like fermions. Labelling $a, b, c, ...$ the components with ordinary femions, and $\alpha, \beta, \gamma, ...$ the exotic ones, the unitary condition in (\ref{unitarity}) can be written in tensor form as

\begin{eqnarray}
\delta _{ij} &=&\left(V_{L,R}^{\ast }\right)_{ij}\left(V_{L,R}\right)_{jk} \nonumber \\
&=&\left(V_{L,R}^{\ast }\right)_{ia}\left(V_{L,R}\right)_{ak}+ \left(V_{L,R}^{\ast }\right)_{i\alpha}\left(V_{L,R}\right)_{\alpha k}.
\end{eqnarray} 
In particular, for the SM components:

\begin{eqnarray}
\delta _{cb} &=&\left(V_{L,R}^{\ast }\right)_{ca}\left(V_{L,R}\right)_{ab}+ \left(V_{L,R}^{\ast }\right)_{c\alpha}\left(V_{L,R}\right)_{\alpha b}.
\label{unitarity-2}
\end{eqnarray} 
Thus, the pure SM submatrix $\left(V_{L,R}\right)_{ab}$ does not satisfy an exact unitary relation, but it is deviated by a small contribution due to new physics from the extra particle content. The relation (\ref{unitarity-2}) is conveniently written as:

\begin{eqnarray}
\left(V_{L,R}^{\ast }\right)_{ca}\left(V_{L,R}\right)_{ab}=\delta _{cb}-\left(V_{L,R}^{\ast }\right)_{c\alpha}\left(V_{L,R}\right)_{\alpha b}.
\label{unitarity-3}
\end{eqnarray}

\subsubsection*{Gauge bosons}

After the symmetry breaking, we obtain from the kinetic Lagrangian in (\ref{eq:Kinetic-Lagrangian}) the charged mass eigenstates 

\begin{eqnarray}
W^{\pm}_{\mu}=\frac{1}{\sqrt{2}}\left(W_{\mu}^1\mp W_{\mu}^2 \right),
\label{gauge-charged}
\end{eqnarray}
with squared mass $M_{\pm}^2=g^2\upsilon ^2/4$, where the electroweak vacuum expectation value $\upsilon = 246$ GeV is defined with the VEV of each scalar doublet as

\begin{eqnarray}
\upsilon=\sqrt{\upsilon _1^2+\upsilon _2^2+\upsilon _3^2}.
\label{electroweak-VEV}
\end{eqnarray}

As for the neutral gauge sector, we obtain in the basis $(W^3_{\mu}, B_{\mu}, Z'_{\mu})$ the following symmetric squared mass matrix:

\begin{eqnarray}
M_0^2=\frac{g^2}{4}
\begin{pmatrix}
\upsilon ^2 & -T_W \upsilon ^2 & | & -\frac{2g_X}{3g}\left(\upsilon ^2 + \upsilon _1 ^2  \right) \\
* & T_W^2 \upsilon ^2 & | & \frac{2g_X}{3g}T_W\left(\upsilon ^2 + \upsilon _1 ^2  \right) \\
 - & - & - & -\ \ \ \ \ \ - \\
* & * & | & \frac{4g_X^2}{9g^2}\left(\upsilon _{\chi} ^2 +\upsilon ^2 + 3 \upsilon _1 ^2  \right)
\end{pmatrix}=
\begin{pmatrix}
A  & | & C \\
 - & -  & - \\
C^T  & | & D
\end{pmatrix},
\label{neutral-gauge-matrix}
\end{eqnarray}
where $T_W=\tan \theta _W$ is the tangent of the Weinberg angle. Taking into account the hierarchy $\upsilon _{\chi} \gg \upsilon$, the above mass matrix can be diagonalized analytically by the recursive expansion method \cite{grimus}. First, according to the block diagonalization shown in Appendix \ref{app:block}, we can reduce the above $3 \times 3$ mass matrix into one $2\times 2$ mass matrix and a heavy mass associated to the $Z'$ boson:

\begin{eqnarray}
a&\approx &A-CD^{-1}C^T=N
\begin{pmatrix}
1 & -T_W \\
-T_W & T_W^2
\end{pmatrix}, \nonumber \\
b&\approx &D= \frac{g_X^2}{9}\left(\upsilon _{\chi} ^2 +\upsilon ^2 + 3 \upsilon _1 ^2  \right),
\label{neutral-gauge-matrix-2}
\end{eqnarray} 
where $N=\frac{1}{4}\left(\upsilon ^2-\frac{\left(\upsilon ^2+\upsilon _1^2\right)^2}{\upsilon _{\chi}^2}\right)$, while the transformation matrix that induces the above block diagonalization is:

\begin{eqnarray}
V=
\begin{pmatrix}
I & F \\
-F^T & I
\end{pmatrix}=
\begin{pmatrix}
1 & 0 & -S_{\theta}C_W \\
0 & 1 & S_{\theta}S_W \\
S_{\theta}C_W & -S_{\theta}S_W & 1
\end{pmatrix},
\label{block-rotation}
\end{eqnarray}
where the sine of the mixing angle $\theta $ has been defined as

\begin{eqnarray}
S_{\theta}=\frac{3}{2}\left(\frac{\upsilon ^2+\upsilon _1^2}{\upsilon _{\chi}^2}\right)\frac{g}{g_X}.
\label{mixing-angle}
\end{eqnarray}
We clarify that in general an additional $Z-Z'$ mixing angle results from the gauge kinetic terms, which can be neglected at a higher scale. This mixing may also arise due to radiative corrections. However, any $Z-Z'$ mixing arised in the model is very restricted by the LEP data, limiting $S_{\theta}$ to small values. In reference \cite{dilepton} the deviations on the Z pole observables due to the mixing angle were evaluated in a $U(1)_X$ model with the same gauge couplings as here, showing allowed mixing angle of the order up to $10^{-4}$
       
Second, the submatrix $a$ in (\ref{neutral-gauge-matrix-2}) has the following mass eigenvalues:

\begin{eqnarray}
m_A^2=0, \ \ \ \ m_Z^2=\frac{g^2}{C_W^2}N,
\label{neutral-masses}
\end{eqnarray}
while the associated rotation matrix is:

\begin{eqnarray}
p=
\begin{pmatrix}
S_W & C_W \\
C_W & -S_W
\end{pmatrix}.
\label{sub-rotation}
\end{eqnarray}
The total rotation into mass eigenstates is the combination of the rotations (\ref{block-rotation}) and (\ref{sub-rotation}),

\begin{eqnarray}
R_0=PV=
\begin{pmatrix}
p & 0 \\
0 & 1
\end{pmatrix}V=
\begin{pmatrix}
S_W & C_W & 0 \\
C_W & -S_W &  S_{\theta} \\
-S_{\theta}C_W &  S_{\theta}S_W & 1
\end{pmatrix},
\label{mass-rotation}
\end{eqnarray}
obtaining the mass eigenstates:

\begin{eqnarray}
\tilde{V}_{\mu}=R_0V_{\mu} \ \ \Rightarrow \ \
\begin{pmatrix}
A_{\mu} \\
Z_{1\mu} \\
Z_{2\mu}
\end{pmatrix}=
R_0
\begin{pmatrix}
W_{\mu}^3 \\
B_{\mu} \\
Z'_{\mu}
\end{pmatrix},
\label{gauge-mass-eigenvectors}
\end{eqnarray}
where $A_{\mu}$ is identified with the photon. We see that in the limit $S_ {\theta}=0$, we obtain $Z_1=Z=C_WW^3-S_WB$ and $Z_2=Z'$, with $Z$ the SM neutral gauge boson. 


\subsubsection*{Neutral currents}

The weak interaction of fermions is contained into the Dirac Lagrangian in (\ref{Dirac-Lagrangian}). First, taking into account the mass eigenstates in (\ref{gauge-charged}) and (\ref{gauge-mass-eigenvectors}), the covariant derivative become

\begin{eqnarray}
D^{\mu }=\partial ^{\mu}-ig\left(W^{\mu+}T_f^{-}+W^{\mu-}T_f^{+}\right)-\tilde{V}^{\mu }_m\left[ig\left(R_0^T\right)_{1m}T_f^3+ig'\frac{Y_f}{2}\left(R_0^T\right)_{2m}+ig_XX_f\left(R_0^T\right)_{3m}\right],
\label{Covariant-eigenmass}
\end{eqnarray}
where $2T_f^{\pm}$ is the combination $(\sigma _1 \pm \sigma _2)$ between the first two Pauli matrices and $2T_f^3$ the third Pauli matrix for fermion fields $f$ doublets of $SU(2)$, while $2T_f^{\pm}=2T_f^{3}=0$ when $f$ are singlets. The terms $(R_0^T)_{nm}$ correspond to the components of the transpose rotation matrix between the neutral weak and mass eigenstates, as defined in (\ref{mass-rotation}), and $\tilde{V}^{\mu }_m$ the corresponding neutral gauge bosons in mass eigenstate, where $\left(\tilde{V}^{\mu }_1, \tilde{V}^{\mu }_2, \tilde{V}^{\mu }_3\right)=\left(A^{\mu }, Z_1^{\mu }, Z_2^{\mu }\right)$. Applying the above covariant derivative into the Dirac Lagrangian (\ref{Dirac-Lagrangian}), we obtain the following neutral gauge interactions:

\begin{eqnarray}
\mathcal{L}_{NC} &=&\frac{g}{2}\left[\overline{f_{Li}}\gamma _{\mu}\tilde{V}^{\mu }_mg_{Lm}^{(f_i)}f_{Li}+\overline{f_{Ri}}\gamma _{\mu}\tilde{V}^{\mu }_mg_{Rm}^{(f_i)}f_{Ri}\right],
\label{Neutral-Current}
\end{eqnarray}
where $g_{L,Rm}^{(f_i)}$ are the electroweak neutral current couplings, defined in general as:

\begin{eqnarray} 
g_{m}^{(f)}&=&\pm \left(R_0^T\right)_{1m}+T_WY_{f}\left(R_0^T\right)_{2m}+\frac{2g_X}{g}X_{f}\left(R_0^T\right)_{3m},
\label{ENC-d}
\end{eqnarray}   
for fermions in doublet representations, where the $\pm$ sign is associated to the upper or lower component of the doublet, and

\begin{eqnarray} 
g_{m}^{(f)}&=&T_WY_{f}\left(R_0^T\right)_{2m}+\frac{2g_X}{g}X_{f}\left(R_0^T\right)_{3m},
\label{ENC-s}
\end{eqnarray} 
for singlets. In particular, for the ordinary SM fermions, labeled with the index $a$, the left-handed couplings are:

\begin{table}
\centering
\renewcommand{\arraystretch}{1.5}
\begin{tabular}{c||c|c|c}\hline
 $f_{La}$ & $g_{L1}^{(f_a)}$ & $g_{L2}^{(f_a)}$ & $g_{L3}^{(f_a)}$ \\ \hline \hline
 $u^1_L$ & \multirow{2}{*}{$\frac{4}{3}S_W$} & $\left(1-\frac{4}{3}S_W^2\right)\frac{1}{C_W}+\frac{2g_X}{3g}S_{\theta}$ & $\left(-1+\frac{4}{3}S_W^2\right)\frac{S_{\theta}}{C_W}+\frac{2g_X}{3g}$ \\ \cline{1-1} \cline{3-4}
$u^{2,3}_L$ & & $\left(1-\frac{4}{3}S_W^2\right)\frac{1}{C_W}$ &  $\left(-1+\frac{4}{3}S_W^2\right)\frac{S_{\theta}}{C_W}$  \\ \hline
 $d^1_L$ & \multirow{2}{*}{$-\frac{2}{3}S_W$} & $\left(-1+\frac{2}{3}S_W^2\right)\frac{1}{C_W}+\frac{2g_X}{3g}S_{\theta}$ & $\left(1-\frac{2}{3}S_W^2\right)\frac{S_{\theta}}{C_W}+\frac{2g_X}{3g}$ \\ \cline{1-1} \cline{3-4}
$d^{2,3}_L$ & & $\left(-1+\frac{2}{3}S_W^2\right)\frac{1}{C_W}$ & $\left(1-\frac{2}{3}S_W^2\right)\frac{S_{\theta}}{C_W}$ \\ \hline
 $e^{e,\mu }_L$ & \multirow{2}{*}{$-2S_W$} & $\left(-1+2S_W^2\right)\frac{1}{C_W}$ & $\left(1-2S_W^2\right)\frac{S_{\theta}}{C_W}$ \\ \cline{1-1} \cline{3-4}
$e^{\tau }_L$ & & $\left(-1+2S_W^2\right)\frac{1}{C_W}-\frac{2g_X}{g}S_{\theta}$ &  $\left(1-2S_W^2\right)\frac{S_{\theta}}{C_W}-\frac{2g_X}{g}$ \\ \hline
 $\nu ^{e,\mu }_L$ & \multirow{2}{*}{$0$} & $\frac{1}{C_W}$ & $\frac{S_{\theta}}{C_W}$  \\ \cline{1-1} \cline{3-4}
$\nu ^{\tau }_L$ & & $\frac{1}{C_W}-\frac{2g_X}{g}S_{\theta}$ & $-\frac{S_{\theta}}{C_W}-\frac{2g_X}{g}$ \\ \hline
\end{tabular}
\caption{Neutral current couplings for the ordinary SM left-handed fermions}
\label{tab:NCL}
\end{table}

\begin{table}
\centering
\renewcommand{\arraystretch}{1.7}
\begin{tabular}{c||c|c|c}\hline
 $f_{Ra}$ & $g_{R1}^{(f_a)}$ & $g_{R2}^{(f_a)}$ & $g_{R3}^{(f_a)}$ \\ \hline \hline
 $u^{1,2,3}_R$ & $\frac{4}{3}S_W$ & $-\frac{4}{3}\left(\frac{S_W^2}{C_W}-\frac{g_X}{g}S_{\theta}\right)$ & $\frac{4}{3}\left(\frac{S_W^2}{C_W}S_{\theta}+\frac{g_X}{g}\right)$ \\ \hline
 $d^{1,2,3}_R$ & $-\frac{2}{3}S_W$ & $\frac{2}{3}\left(\frac{S_W^2}{C_W}-\frac{g_X}{g}S_{\theta}\right)$ & $-\frac{2}{3}\left(\frac{S_W^2}{C_W}S_{\theta}+\frac{g_X}{g}\right)$ \\ \hline
$e^{e,\tau }_R$ & \multirow{2}{*}{$-2S_W$} & $\frac{8}{3}\left(\frac{3}{4}\frac{S_W^2}{C_W}-\frac{g_X}{g}S_{\theta}\right)$ & $-\frac{8}{3}\left(\frac{3}{4}\frac{S_W^2}{C_W}S_{\theta}+\frac{g_X}{g}\right)$ \\ \cline{1-1} \cline{3-4}
$e^{\mu }_R$ & & $\frac{2}{3}\left(3\frac{S_W^2}{C_W}-\frac{g_X}{g}S_{\theta}\right)$ &  $-\frac{2}{3}\left(3\frac{S_W^2}{C_W}S_{\theta}+\frac{g_X}{g}\right)$ \\ \hline
\end{tabular}
\caption{Neutral current couplings for the ordinary SM right-handed fermions}
\label{tab:NCR}
\end{table}



\begin{eqnarray} 
g_{L1}^{(f_a)}&=&2Q_{f_a}S_W, \nonumber \\
g_{L2}^{(f_a)}&=&\frac{1}{C_W}\left(I_3-2Q_{f_a}S_W^2\right)+2X_{f_{La}}\frac{g_X}{g}S_{\theta}, \nonumber \\
g_{L3}^{(f_a)}&=&\frac{1}{C_W}\left(-I_3+2Q_{f_a}S_W^2\right)S_{\theta}+2X_{f_{La}}\frac{g_X}{g}, 
\label{left-ENC}
\end{eqnarray} 
and for right-handed fermions

\begin{eqnarray} 
g_{R1}^{(f_a)}&=&2Q_{f_a}S_W, \nonumber \\
g_{R2}^{(f_a)}&=&-2Q_{f_a}\frac{S_W^2}{C_W}+2X_{f_{Ra}}\frac{g_X}{g}S_{\theta}, \nonumber \\
g_{R3}^{(f_a)}&=&2Q_{f_a}\frac{S_W^2}{C_W}S_{\theta}+2X_{f_{Ra}}\frac{g_X}{g},
\label{right-ENC}
\end{eqnarray} 
where $Q_f$ and $X_f$ are the corresponding electric and $U(1)_X$ charges of the fermion $f$, while $I_3$ is the isospin which is $1$ for the upper components and $-1$ for the lower ones. For future reference, we list explicitly in Tables \ref{tab:NCL} and \ref{tab:NCR} the neutral currents for each flavour fermion. We emphasize that particles such as $e_L^{e,\mu}$, $\nu _L^{e,\mu}$, $u_L^{2,3}$ and $d_L^{2,3}$ are devoid of couplings with $g_X$, which is a consequence of their zero $U(1)_X$ charge.  

For the newly fermions, labeled with the index $\alpha $, both the left-handed and right-handed are singlets of $SU(2)_L$. Thus, the neutral current couplings are: 


\begin{eqnarray} 
g_{L,Rm}^{(f_{\alpha})}&=&T_WY_{f_{L,R\alpha}}\left(R_0^T\right)_{2m}+\frac{2g_X}{g}X_{f_{L,R\alpha}}\left(R_0^T\right)_{3m},
\label{extra-ENC}
\end{eqnarray} 
which are listed in Tables \ref{tab:NCLnew} and \ref{tab:NCRnew} for each flavour of this sector.

On the other hand, according to (\ref{fermion-mass-basis}), the fermion fields must be also rotated into a mass eigenstate basis. By labeling $\tilde{f}_i$ each component of the mass basis $\mathbf{\tilde{f}} $ and $f_i$ the corresponding in weak basis, the transformation (\ref{fermion-mass-basis}) are written in components as:

\begin{eqnarray}
\tilde{f}_{L,Ri}=\left(V_{L,R}^{f\dag}\right)_{ij}f_{L,Rj}.
\end{eqnarray}
Thus, the neutral current Lagrangian (\ref{Neutral-Current}) in full mass eigenstates is:

\begin{eqnarray}
\mathcal{L}_{NC} &=&\frac{g}{2}\left[ \overline{\tilde f_{Li}}\gamma _{\mu}\tilde{V}^{\mu }_m\left(V_L^{f\dag}\right)_{ij}g_{Lm}^{(f_j)}\left(V_L^{f}\right)_{jk}\tilde f_{Lk} \right. \nonumber \\
&&\left. + \overline{\tilde f_{Ri}}\gamma _{\mu}\tilde{V}^{\mu }_m\left(V_R^{f\dag}\right)_{ij}g_{Rm}^{(f_j)}\left(V_R^{f}\right)_{jk}\tilde f_{Rk}\right].
\label{Neutral-Current-mass}
\end{eqnarray}
In mass eigenstates, the neutral current couplings transform through the fermionic bi-unitary matrices:

\begin{table}
\centering
\renewcommand{\arraystretch}{1.7}
\begin{tabular}{c||c|c|c}\hline
 $f_{L\alpha}$ & $g_{L1}^{(f_{\alpha})}$ & $g_{L2}^{(f_{\alpha})}$ & $g_{L3}^{(f_{\alpha})}$ \\ \hline \hline
 $\mathcal{T}_L$ & $\frac{4}{3}S_W$ & $-\frac{2}{3}\left(2\frac{S_W^2}{C_W}-\frac{g_X}{g}S_{\theta}\right)$ & $\frac{2}{3}\left(2\frac{S_W^2}{C_W}S_{\theta}+\frac{g_X}{g}\right)$ \\ \hline
$\mathcal{J}^{1,2}_L$ & $-\frac{2}{3}S_W$ & $\frac{2}{3}\frac{S_W^2}{C_W}$ &  $-\frac{2}{3}\frac{S_W^2}{C_W}S_{\theta}$  \\ \hline
 $\mathcal{E}^{1}_L$ & \multirow{2}{*}{$-2S_W$} & $2\left(\frac{S_W^2}{C_W}-\frac{g_X}{g}S_{\theta}\right)$ &  $-2\left(\frac{S_W^2}{C_W}S_{\theta }+\frac{g_X}{g}\right)$ \\ \cline{1-1} \cline{3-4}
 $\mathcal{E}^{2}_L$ &  & $\frac{4}{3}\left(\frac{3}{2}\frac{S_W^2}{C_W}-\frac{g_X}{g}S_{\theta}\right)$ &  $-\frac{4}{3}\left(\frac{3}{2}\frac{S_W^2}{C_W}S_{\theta }+\frac{g_X}{g}\right)$ \\ \hline
\end{tabular}
\caption{Neutral current couplings for the newly left-handed fermions}
\label{tab:NCLnew}
\end{table}

\begin{table}
\centering
\renewcommand{\arraystretch}{1.7}
\begin{tabular}{c||c|c|c}\hline
 $f_{R\alpha}$ & $g_{R1}^{(f_{\alpha})}$ & $g_{R2}^{(f_{\alpha})}$ & $g_{R3}^{(f_{\alpha})}$ \\ \hline \hline
 $\mathcal{T}_R$ & $\frac{4}{3}S_W$ & $-\frac{4}{3}\left(\frac{S_W^2}{C_W}-\frac{g_X}{g}S_{\theta}\right)$ & $\frac{4}{3}\left(\frac{S_W^2}{C_W}S_{\theta}+\frac{g_X}{g}\right)$ \\ \hline
$\mathcal{J}^{1,2}_R$ & $-\frac{2}{3}S_W$ & $\frac{2}{3}\left(\frac{S_W^2}{C_W}-\frac{g_X}{g}S_{\theta }\right)$ &  $-\frac{2}{3}\left(\frac{S_W^2}{C_W}S_{\theta}+\frac{g_X}{g}\right)$  \\ \hline
 $\mathcal{E}^{1}_R$ & \multirow{2}{*}{$-2S_W$} & $\frac{4}{3}\left(\frac{3}{2}\frac{S_W^2}{C_W}-\frac{g_X}{g}S_{\theta}\right)$ &  $-\frac{4}{3}\left(\frac{3}{2}\frac{S_W^2}{C_W}S_{\theta }+\frac{g_X}{g}\right)$ \\ \cline{1-1} \cline{3-4}
 $\mathcal{E}^{2}_R$ &  & $2\left(\frac{S_W^2}{C_W}-\frac{g_X}{g}S_{\theta}\right)$ &  $-2\left(\frac{S_W^2}{C_W}S_{\theta }+\frac{g_X}{g}\right)$ \\ \hline
$\nu ^{1, 2, 3}_R$ & $0$ & $\frac{2}{3}\frac{g_X}{g}S_{\theta}$ & $\frac{2}{3}\frac{g_X}{g}$ \\ \hline
$\mathcal{N}^{1, 2, 3}_R$ & $0$ & $0$ & $0$ \\ \hline
\end{tabular}
\caption{Neutral current couplings for the newly right-handed fermions}
\label{tab:NCRnew}
\end{table}

\begin{eqnarray}
g_{L,Rm}^{(f_j)} \longrightarrow \tilde g_{L,Rm}^{(ik)}= \left(V_{L,R}^{f\dag}\right)_{ij}g_{L,Rm}^{(f_j)}\left(V_{L,R}^{f}\right)_{jk},
\label{NC-mass}
\end{eqnarray}
so, the neutral Lagrangian (\ref{Neutral-Current-mass}) become:

\begin{eqnarray}
\mathcal{L}_{NC} &=&\frac{g}{2}\left[ \overline{\tilde f_{Li}}\gamma _{\mu}\tilde{V}^{\mu }_m\tilde g_{Lm}^{(jk)}\tilde f_{Lk} + \overline{\tilde f_{Ri}}\gamma _{\mu}\tilde{V}^{\mu }_m\tilde g_{Rm}^{(jk)}\tilde f_{Rk}\right].
\label{Neutral-Current-mass-2}
\end{eqnarray}
In general, as shown in Tables \ref{tab:NCL} - \ref{tab:NCRnew}, thera are couplings that are family dependent. For these cases, the neutral couplings $\tilde g_{L,R}^{(ik)}$ are non-diagonal, producing FCNC processes, such as in the dilepton B decay. For the family universal couplings, due to the unitary constraint in (\ref{unitarity}), the neutral couplings become diagonal, $\tilde g_{L,R}^{(ik)}=g_{L,R}^{(f_j)}\delta _{ik}$, which only produce flavour conservative neutral currents.

\section{B decay}

The process $B^+\rightarrow K^+\ell^+\ell^-$ for chaged leptons $\ell ^{\pm}$ is due to $b\rightarrow s\ell^+\ell^-$ transitions. In the model, this process can be induced at tree level through the neutral weak bosons $Z_1$ and $Z_2$ as shown in Figure \ref{fig:Bs-decay}. 

\subsection{Fundamental couplings}

\subsubsection*{The $b-s-Z_{1(2)}$ coupling}

\begin{figure}
\centering
\includegraphics[scale=0.25]{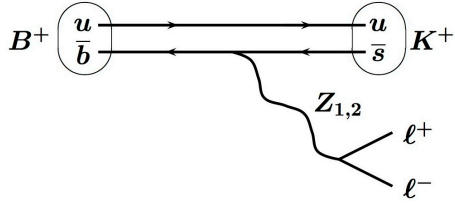}\vspace{-0.5cm}
\caption{Decay $B^+\rightarrow K^+\ell^+\ell^-$ through neutral gauge bosons $Z_{1,2}$.}
\label{fig:Bs-decay}
\end{figure}

First, according to the neutral current Lagrangian in (\ref{Neutral-Current-mass-2}), the FCNC transition $b \rightarrow s$ in the first vertex of Figure \ref{fig:Bs-decay}, is described by the Lagrangian:

\begin{eqnarray}
\mathcal{L}_{sb} &=&\frac{g}{2}\left[ \overline{s_{L}}\gamma _{\mu}\left(Z_1^{\mu }\tilde g_{L2}^{(23)}+Z_2^{\mu }\tilde g_{L3}^{(23)} \right)b_{L} + \overline{ s_{R}}\gamma _{\mu}\left(Z_1^{\mu }\tilde g_{R2}^{(23)}+Z_2^{\mu }\tilde g_{R3}^{(23)} \right)b_{R}\right]+\text{H.c.},
\label{bs-lagrangian}
\end{eqnarray}
where:

\begin{eqnarray}
\tilde g_{L,Rm}^{(23)}= \left(V_{L,R}^{D\dag}\right)_{2j}g_{L,Rm}^{(D_j)}\left(V_{L,R}^{D}\right)_{j3},
\end{eqnarray}
with $D_j=(d_1,d_2,d_3, \mathcal{J}_1,\mathcal{J}_2)$. Separating the ordinary fermions $D_a=(d_1,d_2,d_3)$ from the new ones $D_{\alpha }=(\mathcal{J}_1,\mathcal{J}_2)$, we can write the above coupling as:

\begin{eqnarray}
\tilde g_{L,Rm}^{(23)}= \left(V_{L,R}^{D\dag}\right)_{2a}g_{L,Rm}^{(D_a)}\left(V_{L,R}^{D}\right)_{a3}
+\left(V_{L,R}^{D\dag}\right)_{2\alpha}g_{L,Rm}^{(D_{\alpha })}\left(V_{L,R}^{D}\right)_{\alpha 3}.
\label{NC-32}
\end{eqnarray}
Taking into account that according to tables \ref{tab:NCL} - \ref{tab:NCRnew} for the down-type sector, only the left-handed ordinary down quarks exhibits family dependence, then the left-handed couplings in (\ref{NC-32}) expands as:

\begin{eqnarray}
\tilde g_{Lm}^{(23)}&=&g_{Lm}^{(d_1)}\left(V_{L}^{D\dag}\right)_{21}\left(V_{L,R}^{D}\right)_{13}+
g_{Lm}^{(d_{2,3})}\left[\left(V_{L}^{D\dag}\right)_{22}\left(V_{L,R}^{D}\right)_{23}
+\left(V_{L,R}^{D\dag}\right)_{23}\left(V_{L,R}^{D}\right)_{33}\right] \nonumber \\
&&+ g_{Lm}^{(D_{\alpha})}\left(V_{L}^{D\dag}\right)_{2\alpha}\left(V_{L,R}^{D}\right)_{\alpha 3},
\label{NC-32left}
\end{eqnarray}
while the right-handed couplings (family universal) cancel out,

\begin{eqnarray}
\tilde g_{Rm}^{(23)}&=&0.
\label{NC-32right}
\end{eqnarray}
From the unitary constraint (\ref{unitarity-3}), we find the following relation:

\begin{eqnarray}
\left(V_{L}^{D\dag}\right)_{22}\left(V_{L}^{D}\right)_{23}+\left(V_{L}^{D\dag}\right)_{23}\left(V_{L}^{D}\right)_{33}
=-\left(V_{L}^{D\dag}\right)_{21}\left(V_{L}^{D}\right)_{13}-\left(V_{L}^{D\dag}\right)_{2\alpha}\left(V_{L}^{D}\right)_{\alpha 3},
\end{eqnarray} 
which, after replacing in (\ref{NC-32left}), we obtain:

\begin{eqnarray}
\tilde g_{Lm}^{(23)}&=&\left(V_{L}^{D\dag}\right)_{21}\left(V_{L}^{D}\right)_{13}\left[g_{Lm}^{(d_1)}-g_{Lm}^{(d_{2,3})}\right]+\left(V_{L}^{D\dag}\right)_{2\alpha }\left(V_{L}^{D}\right)_{\alpha 3}\left[g_{Lm}^{(D_{\alpha })}-g_{Lm}^{(d_{2,3})}\right].
\label{NC-32left2}
\end{eqnarray}
We see that in a family universal scenario, where the coupling $g_{Lm}^{(d_{2,3})}$ would be the same as for $g_{Lm}^{(d_{1})}$ and $g_{Lm}^{(D_{\alpha})}$, the above coupling cancel out, suppressing the FCNC transition $b\rightarrow s$. However, the model distinguish these couplings, according to the family index. Specifically, using the values from Table \ref{tab:NCL}, we obtain the left-handed neutral couplings for the $b-s$ interaction shown in the first row from Table \ref{tab:NCBLdecay}. 



\subsubsection*{The $e^+(\mu ^+)-e^-(\mu ^-) -Z_{1(2)}$ coupling}

On the other hand, the neutral coupling for the decays $Z_{1,2}\rightarrow \ell^{+}_{a}\ell^{-}_{a}$ in the second vertex from Figure \ref{fig:Bs-decay} for $\ell _{a}= e$ and $\mu $, is described by:

\begin{eqnarray}
\mathcal{L}_{\ell} &=&\frac{g}{2}\left[ \overline{\ell_{La}}\gamma _{\mu}\left(Z_1^{\mu }\tilde g_{L2}^{(aa)}+Z_2^{\mu }\tilde g_{L3}^{(aa)} \right)\ell _{La} + \overline{ \ell_{Ra}}\gamma _{\mu}\left(Z_1^{\mu }\tilde g_{R2}^{(aa)}+Z_2^{\mu }\tilde g_{R3}^{(aa)} \right)\ell_{Ra}\right],
\label{aa-lagrangian}
\end{eqnarray}
with

\begin{eqnarray}
\tilde g_{L,Rm}^{(aa)}= \left(V_{L,R}^{E\dag}\right)_{aj}g_{L,Rm}^{(E_j)}\left(V_{L,R}^{E}\right)_{ja},
\end{eqnarray}
for $E_j=(e^{e},e^{\mu }, e^{\tau}, \mathcal{E}_1, \mathcal{E}_2)$. By using the unitary constraint, we obtain for the left-handed couplings of the charged leptons:

\begin{eqnarray}
\tilde g_{Lm}^{(aa)}=g_{Lm}^{(e^e,e^{\mu })}&+&\left|\left(V_{L}^{E}\right)_{3a}\right|^2\left[g_{Lm}^{(e^{\tau })}-g_{Lm}^{(e^e,e^{\mu })}\right] \nonumber \\
&+&\left|\left(V_{L}^{E}\right)_{\alpha a}\right|^2\left[g_{Lm}^{(E_{\alpha })}-g_{Lm}^{(e^e,e^{\mu })}\right],
\end{eqnarray}
and for the right-handed ones, we obtain:

\begin{eqnarray}
\tilde g_{Rm}^{(aa)}=g_{Rm}^{(e^e,e^{\tau })}&+&\left|\left(V_{R}^{E}\right)_{2a}\right|^2\left[g_{Rm}^{(e^{\mu })}-g_{Rm}^{(e^e,e^{\tau })}\right] \nonumber \\
&+&\left|\left(V_{R}^{E}\right)_{\alpha a}\right|^2\left[g_{Rm}^{(E_{\alpha })}-g_{Rm}^{(e^e,e^{\tau })}\right].
\end{eqnarray}
In this case, we see that the first term of the above equations do not depend on the flavour number $a$ (it is the same for $e^{\pm}$ as for $\mu ^{\pm}$). However, the subsequent terms depends explicitly from the $ia$ components of the fermionic bi-unitary matrices, due to the nonuniversality of the neutral couplings. Since, in general, each component of the matrices are different, we will obtain a distinction between the couplings to electrons and to muons. As a consequence, the ratio of the branching of the $B^+\rightarrow K^+e^+e^-$ and $B^+\rightarrow K^+\mu^+\mu^-$ deviates from one, as suggests the LHCb data. Again, using the values from Tables \ref{tab:NCL} - \ref{tab:NCRnew} for the charged leptons, we obtain the neutral couplings for $Z_{1,2}\rightarrow e^{\pm} (\mu ^{\pm})$ in tables \ref{tab:NCBLdecay} and \ref{tab:NCBRdecay}, for left- and right-handed leptons, respectively.


\subsection{Effective operators}

\begin{table}
\centering
\renewcommand{\arraystretch}{1.7}
\begin{tabular}{c|| c |c}\hline
 $\overline{f_a}f_b$ & $\tilde g_{L2}^{(ab)}$ & $\tilde g_{L3}^{(ab)}$  \\ \hline \hline
 $\overline{s}b$ & {\tiny $\frac{2}{3}\frac{g_X}{g}\left(V_{L}^{D\dag}\right)_{21}\left(V_{L}^{D}\right)_{13}S_{\theta}+\frac{1}{C_W}\left(V_{L}^{D\dag}\right)_{2\alpha }\left(V_{L}^{D}\right)_{\alpha 3}$} & {\tiny $\frac{2}{3}\frac{g_X}{g}\left(V_{L}^{D\dag}\right)_{21}\left(V_{L}^{D}\right)_{13}-\frac{1}{C_W}\left(V_{L}^{D\dag}\right)_{2\alpha }\left(V_{L}^{D}\right)_{\alpha 3}S_{\theta}$}  \\ \hline
\multirow{2}{*}{$e^+e^-$} & {\tiny $\left(-1+2S_W^2\right)\frac{1}{C_W}-2\frac{g_X}{g}\left|\left(V_{L}^{E}\right)_{31}\right|^2S_{\theta}$} & {\tiny $\left(1-2S_W^2\right)\frac{S_{\theta}}{C_W}-2\frac{g_X}{g}\left|\left(V_{L}^{E}\right)_{31}\right|^2$}  \\ 
    & {\tiny $+\left(\frac{1}{C_W}-2\frac{g_X}{g}S_{\theta }\right)\left|\left(V_{L}^{E}\right)_{41}\right|^2+\left(\frac{1}{C_W}-\frac{4}{3}\frac{g_X}{g}S_{\theta }\right)\left|\left(V_{L}^{E}\right)_{51}\right|^2$} & {\tiny $-\left(\frac{S_{\theta }}{C_W}+2\frac{g_X}{g}\right)\left|\left(V_{L}^{E}\right)_{41}\right|^2-\left(\frac{S_{\theta }}{C_W}+\frac{4}{3}\frac{g_X}{g}\right)\left|\left(V_{L}^{E}\right)_{51}\right|^2$} \\ \hline
 \multirow{2}{*}{$\mu ^+\mu ^-$} & {\tiny $\left(-1+2S_W^2\right)\frac{1}{C_W}-2\frac{g_X}{g}\left|\left(V_{L}^{E}\right)_{32}\right|^2S_{\theta}$} & {\tiny $\left(1-2S_W^2\right)\frac{S_{\theta}}{C_W}-2\frac{g_X}{g}\left|\left(V_{L}^{E}\right)_{32}\right|^2$}   \\ 
    & {\tiny $+\left(\frac{1}{C_W}-2\frac{g_X}{g}S_{\theta }\right)\left|\left(V_{L}^{E}\right)_{42}\right|^2+\left(\frac{1}{C_W}-\frac{4}{3}\frac{g_X}{g}S_{\theta }\right)\left|\left(V_{L}^{E}\right)_{52}\right|^2$} & {\tiny $-\left(\frac{S_{\theta }}{C_W}+2\frac{g_X}{g}\right)\left|\left(V_{L}^{E}\right)_{42}\right|^2-\left(\frac{S_{\theta }}{C_W}+\frac{4}{3}\frac{g_X}{g}\right)\left|\left(V_{L}^{E}\right)_{52}\right|^2$} \\ \hline
\end{tabular}
\caption{Neutral current couplings for the left-handed fermions $b-s$, $e^{\pm}$ and $\mu ^{\pm}$}
\label{tab:NCBLdecay}
\end{table}

\begin{table}
\centering
\renewcommand{\arraystretch}{1.7}
\begin{tabular}{c|| c |c}\hline
 $\overline{f_a}f_b$ & $\tilde g_{R2}^{(ab)}$ & $\tilde g_{R3}^{(ab)}$  \\ \hline \hline
 $\overline{s}b$ & 0 & 0 \\ \hline
\multirow{2}{*}{$e^+e^-$} & {\tiny $\frac{8}{3}\left(\frac{3}{4}\frac{S_W^2}{C_W}-\frac{g_X}{g}S_{\theta}\right)+2\frac{g_X}{g}\left|\left(V_{R}^{E}\right)_{21}\right|^2S_{\theta}$} & {\tiny $-\frac{8}{3}\left(\frac{3}{4}\frac{S_W^2}{C_W}S_{\theta}+\frac{g_X}{g}\right)+2\frac{g_X}{g}\left|\left(V_{R}^{E}\right)_{21}\right|^2$}  \\ 
    & {\tiny $+\frac{4}{3}\frac{g_X}{g}\left|\left(V_{R}^{E}\right)_{41}\right|^2S_{\theta }+\frac{2}{3}\frac{g_X}{g}\left|\left(V_{R}^{E}\right)_{51}\right|^2S_{\theta }$} & {\tiny $\frac{4}{3}\frac{g_X}{g}\left|\left(V_{R}^{E}\right)_{41}\right|^2+\frac{2}{3}\frac{g_X}{g}\left|\left(V_{R}^{E}\right)_{51}\right|^2$} \\ \hline
 \multirow{2}{*}{$\mu ^+\mu ^-$} & {\tiny $\frac{8}{3}\left(\frac{3}{4}\frac{S_W^2}{C_W}-\frac{g_X}{g}S_{\theta}\right)+2\frac{g_X}{g}\left|\left(V_{R}^{E}\right)_{22}\right|^2S_{\theta}$} & {\tiny $-\frac{8}{3}\left(\frac{3}{4}\frac{S_W^2}{C_W}S_{\theta}+\frac{g_X}{g}\right)+2\frac{g_X}{g}\left|\left(V_{R}^{E}\right)_{22}\right|^2$}  \\ 
     & {\tiny $+\frac{4}{3}\frac{g_X}{g}\left|\left(V_{R}^{E}\right)_{42}\right|^2S_{\theta }+\frac{2}{3}\frac{g_X}{g}\left|\left(V_{R}^{E}\right)_{52}\right|^2S_{\theta }$} & {\tiny $\frac{4}{3}\frac{g_X}{g}\left|\left(V_{R}^{E}\right)_{42}\right|^2+\frac{2}{3}\frac{g_X}{g}\left|\left(V_{R}^{E}\right)_{52}\right|^2$} \\ \hline
\end{tabular}
\caption{Neutral current couplings for the right-handed fermions $b-s$, $e^{\pm}$ and $\mu ^{\pm}$}
\label{tab:NCBRdecay}
\end{table}

From the neutral Lagrangians in (\ref{bs-lagrangian}) and (\ref{aa-lagrangian}), we obtain the matrix element for the $b\rightarrow s\ell_{a}^+\ell _a^-$ proccess:

\begin{eqnarray}
i\mathcal{M}_{fi}&=&-\frac{ig^2}{4}\left[\overline{u_{s}}\gamma _{\mu}\left(\tilde g_{Lm}^{(23)}L\right)u_{b}\right]D^{\mu \nu}\left[\overline{u_{a}}\gamma _{\nu}\left(\tilde g_{Lm}^{(aa)}L+\tilde g_{Rm}^{(aa)}R\right)v_a\right],
\end{eqnarray}
where $u_{s,b,a}$ are the wave functions of the fermions $s, b$ and $\ell _a$, respectively, and  $v_a$ of anti-leptons $\overline{\ell }_a$, while $D^{\mu \nu}$ is the propagator of the intermediary gauge bosons, defined in the Feynman gauge as:

\begin{eqnarray}
D^{\mu \nu}=\frac{-ig^{\mu \nu}}{q^2-M_{Z_m}^2}.
\end{eqnarray}
At low energies, the momentum transfer through the intermediary particles is negligible in relation to their masses. Thus, the above matrix element become:

\begin{eqnarray}
i\mathcal{M}_{fi}&\approx&-\frac{ig^2}{4M_{Z_m}^2}\overline{u_{s}}\gamma _{\mu}\left(\tilde g_{Lm}^{(23)}L\right)u_{b}\overline{u_{a}}\gamma ^{\mu}\left(\tilde g_{Lm}^{(aa)}L+\tilde g_{Rm}^{(aa)}R\right)v_a.
\end{eqnarray}
The above matrix element can be derived from the following effective Hamiltonian:

\begin{eqnarray}
\mathcal{H}_{\text{eff}}^{\text{NP}}=\frac{g^2}{4M_{Z_m}^2}\left[\overline{s}\left(\tilde g_{Lm}^{(23)}\gamma _{\mu}L\right)b\right]\left[\overline{\ell_{a}}\gamma ^{\mu}\left(\tilde g_{Lm}^{(aa)}L+\tilde g_{Rm}^{(aa)}R\right)\ell_a\right]+\text{H.c},
\label{NP-effective}
\end{eqnarray}
where NP is the label for new non-SM physics, which affect the ordinary SM contribution, described by the Wilson operators through the effective Hamiltonian \cite{SM-effective, SM-effective-2, SM-effective-3}:

\begin{eqnarray}
\mathcal{H}_{\text{eff}}^{\text{SM}}=-\frac{4G_F}{\sqrt{2}}V_{tb}V^{\ast}_{ts}\sum _i\left[C_i^{SM}\mathcal{O}_i+C_i^{'SM}\mathcal{O'}_i\right]+\text{H.c.},
\label{SM-effective}
\end{eqnarray}
where the dominant Wilson coefficients are $C_{i}^{SM}=C_{9,10}^{SM}$, with

\begin{eqnarray}
\mathcal{O}_9&=&\frac{\alpha _{\text{em}}}{4\pi }\left[\overline{s}\gamma _{\mu }Lb\right]\left[\overline{\ell _a}\gamma ^{\mu }\ell _{a}\right], \nonumber \\
\mathcal{O}_{10}&=&\frac{\alpha _{\text{em}}}{4\pi }\left[\overline{s}\gamma _{\mu }Lb\right]\left[\overline{\ell _a}\gamma ^{\mu }\gamma _5\ell _{a}\right].
\label{wilson-operators}
\end{eqnarray}
Putting together both Hamiltonians, equations (\ref{NP-effective}) and (\ref{SM-effective}), and taking the approximated value of 

\begin{eqnarray}
\frac{G_F\alpha _{\text{em}}}{\sqrt{2}\pi}V_{tb}V^{\ast}_{ts} \approx \frac{1}{\left(36 \text{TeV}\right)^2},
\end{eqnarray}
we obtain the total effective Hamiltonian:

\begin{eqnarray}
\mathcal{H}_{\text{eff}}&=&\mathcal{H}_{\text{eff}}^{\text{SM}}+\mathcal{H}_{\text{eff}}^{\text{NP}} \nonumber \\
&=&-\frac{1}{\left(36 \text{TeV}\right)^2}\left[C_9^{SM}-\frac{g^2\left(36 \text{TeV}\right)^2}{8M_{Z_m}^2}\tilde g_{Lm}^{(23)}\left(\tilde g_{Lm}^{(aa)}+\tilde g_{Rm}^{(aa)}\right)\right]\left(\overline{s}\gamma _{\mu }Lb\right)\left(\overline{\ell _a}\gamma ^{\mu }\ell _{a}\right) \nonumber \noindent \\
&&-\frac{1}{\left(36 \text{TeV}\right)^2}\left[C_{10}^{SM}+\frac{g^2\left(36 \text{TeV}\right)^2}{8M_{Z_m}^2}\tilde g_{Lm}^{(23)}\left(\tilde g_{Lm}^{(aa)}-\tilde g_{Rm}^{(aa)}\right)\right]\left(\overline{s}\gamma _{\mu }Lb\right)\left(\overline{\ell _a}\gamma ^{\mu }\gamma _5 \ell _{a}\right),
\label{total-effective}
\end{eqnarray}
from where we identify the total Wilson coefficients:

\begin{eqnarray}
C_9^{(a)}=C_9^{SM}+ C_{9}^{NP(a)}, \ \ \ \ \ \ C_{10}^{(a)}= C_{10}^{SM}+ C_{10}^{NP(a)},
\label{wilson-expanded}
\end{eqnarray}
with:

\begin{eqnarray}
C_{9}^{NP(a)}=-\frac{g^2\left(36 \text{TeV}\right)^2}{8M_{Z_m}^2}\tilde g_{Lm}^{(23)}\left(\tilde g_{Lm}^{(aa)}+\tilde g_{Rm}^{(aa)}\right) \label{NP-WILSON-a}
 \noindent \\
C_{10}^{NP(a)}=\frac{g^2\left(36 \text{TeV}\right)^2}{8M_{Z_m}^2}\tilde g_{Lm}^{(23)}\left(\tilde g_{Lm}^{(aa)}-\tilde g_{Rm}^{(aa)}\right),
\label{NP-WILSON}
\end{eqnarray}
where a sum over repeated indices $m=\{1,2\}$ is implied in the right terms. For the SM contributions, we use the values $C_9^{SM}\approx -C_{10}^{SM}\approx 4.1$ \cite{SM-effective-3}.

\subsection{$e-\mu $ relative branching ratio}

The LHCb collaboration recorded a measurement of the ratio of the branching fractions of $B^+\rightarrow K^+\mu^+\mu^-$ and $B^+\rightarrow K^+e^+e^-$ decay, which is given by:

\begin{eqnarray}
R_K=\frac{\int _{q_{\text{min}}^2} ^{q_{\text{max}}^2}\frac{d\Gamma [B^+\rightarrow K^+\mu^+\mu^-]}{dq^2}dq^2}{\int _{q_{\text{min}}^2} ^{q_{\text{max}}^2}\frac{d\Gamma [B^+\rightarrow K^+e^+e^-]}{dq^2}dq^2}, 
\end{eqnarray} 
within the dilepton invariant mass squared range $1 < q^2 < 6$ GeV$^2/c^4$. In terms of the Wilson coefficients, $R_K$ is \cite{SM-effective-4}:

\begin{eqnarray}
R_K=\frac{\left|C_9^{(\mu )}\right|^2+\left|C_{10}^{(\mu )}\right|^2}{\left|C_9^{(e)}\right|^2+\left|C_{10}^{(e)}\right|^2}.
\label{Rk}
\end{eqnarray}
By expanding the coefficients in SM and NP contributions according to (\ref{wilson-expanded}), and taking into accout the lepton universality of the SM, we obtain:

\begin{eqnarray}
R_K=\frac{\left|C_9^{SM}+C_{9}^{NP(\mu)}\right|^2+\left|C_{10}^{SM}+ C_{10}^{NP(\mu )}\right|^2}{\left|C_9^{SM}+C_{9}^{NP(e )}\right|^2+\left|C_{10}^{SM}+ C_{10}^{NP(e)}\right|^2}.
\label{Rk-2}
\end{eqnarray}

By assuming that the above expression corresponds to the experimentally measured, we can fit the free parameters of the model according to the reported value \cite{lhcb}

\begin{eqnarray}
R_K=0.745^{+0.090}_{-0.074}\pm 0.036.
\label{lhcb-ratio}
\end{eqnarray}

The free parameters are classified into two categories. First, the gauge parameters, corresponding to the $Z'$ gauge boson mass, the gauge coupling constant of the $U(1)_X$ symmetry, and the $Z-Z'$ mixing angle: $(M_{Z'}, g_X, S_{\theta })$. Second, the fermion parameters which arise from the biunitary transformations that rotate the fermion flavours into mass states, according to (\ref{fermion-mass-basis}), and that depend from the Yukawa couplings and the VEVs of the Higgs fields. By using the scheme shown in reference \cite{1s3d-nonuniversal}, these matrices can be parameterized as functions of mixing angles. After some simplifications, as shown in appendix \ref{app:biunitary}, we are left with six free parameters: two ratios of Yukawa couplings, $r_{\mathcal{J}}=h_{\mathcal{J}}/h_u$ and $r_{\mathcal{E}}=h_{\mathcal{E}}/h_u$, where $h_{\mathcal{J}, \mathcal{E}}$ are the couplings of the extra charged fermions shown in the matrices in equations (\ref{eq:Down-mass-matrix-2}) and (\ref{eq:Electron-mass-matrix-2}), while $h_u$ is the coupling of the ordinary up-type quarks according to (\ref{app:up-matrices-2}), the two masses $m_J$ and $m_E$, corresponding to the new down-type quarks and charged leptons, and two mixing angles from the left- and right-handed charged leptons, $\theta _{13}^{E_{L}}$ and $\theta _{25}^{E_{R}}$, which we express through their tangents $t_{13}^{E_L}$ and $t_{25}^{E_R}$. All other mixing angles can be written as function of these two angles, as shown in equation (\ref{leptonmixing-natural}). In particular, as shown in tables \ref{tab:NCBLdecay} and \ref{tab:NCBRdecay}, the neutral current couplings depends on the $ij=2a, 3a, 4a$ and $5a$ bi-unitary components with $a=1$ for electrons and $2$ for muons. Explicitly these components can be fully written as functions of $\theta _{13}^{E_{L}}$ and $\theta _{25}^{E_{R}}$, as shown in equations (\ref{lepton-left-biuni}) and (\ref{lepton-right-biuni}).

Thus, the space of parameters is reduced to 9 variables: $(M_{Z'},m_J, m_E, g_X, r_{\mathcal{J}}, r_{\mathcal{E}}, S_{\theta}, t_{13}^{E_{L}},$ $t_{25}^{E_{R}})$. However, some of these parameters are constrained from theoretical conditions and other experimental observables. For example, the mass $M_{Z'}$ has lower limits from direct detection in colliders. Experiments at LHC collected data at $\sqrt{s}=13$ TeV for new resonances in dielectron and dimuon final states, where lower limits on $M_{Z'}$ between $3.5$ TeV and $4.5$ TeV at 36.1 fb$^{-1}$ by the ATLAS collaboration, and $3.5$ TeV and $4$ TeV at 12.4 fb$^{-1}$ by CMS are reported \cite{z2limits}. We take the lowest experimental limit of 3.5 TeV. Also, in models with extra gauge neutral bosons, the $Z-Z'$ mixing angle is suppresed as the inverse of the squared $Z'$-mass and by electroweak observables, to values up to $\sim 10^{-3}$, which has a negligible effect on the total branching decays. Thus, for simplicity, we ignore this mixing and take $S_{\theta}=0$. The coupling $g_X$ is constrained by $Z'$ production limits. For example, in some models with the same gauge couplings as the model proposed here, limits on dilepton events $pp\rightarrow Z' \rightarrow \ell \ell$ at LHC allow values as large as $g_X\approx 0.4$ \cite{extra-u1, dilepton}. Search for extra fermions can change according to specific model-dependent assumptions \cite{particle data}. We use a safe scenary with mass values around the TeV scale. Finally, we assume one common Yukawa ratio $r_h= r_{\mathcal{J}} = r_{\mathcal{E}}$.

 \begin{figure}
\centering
\includegraphics[scale=0.28]{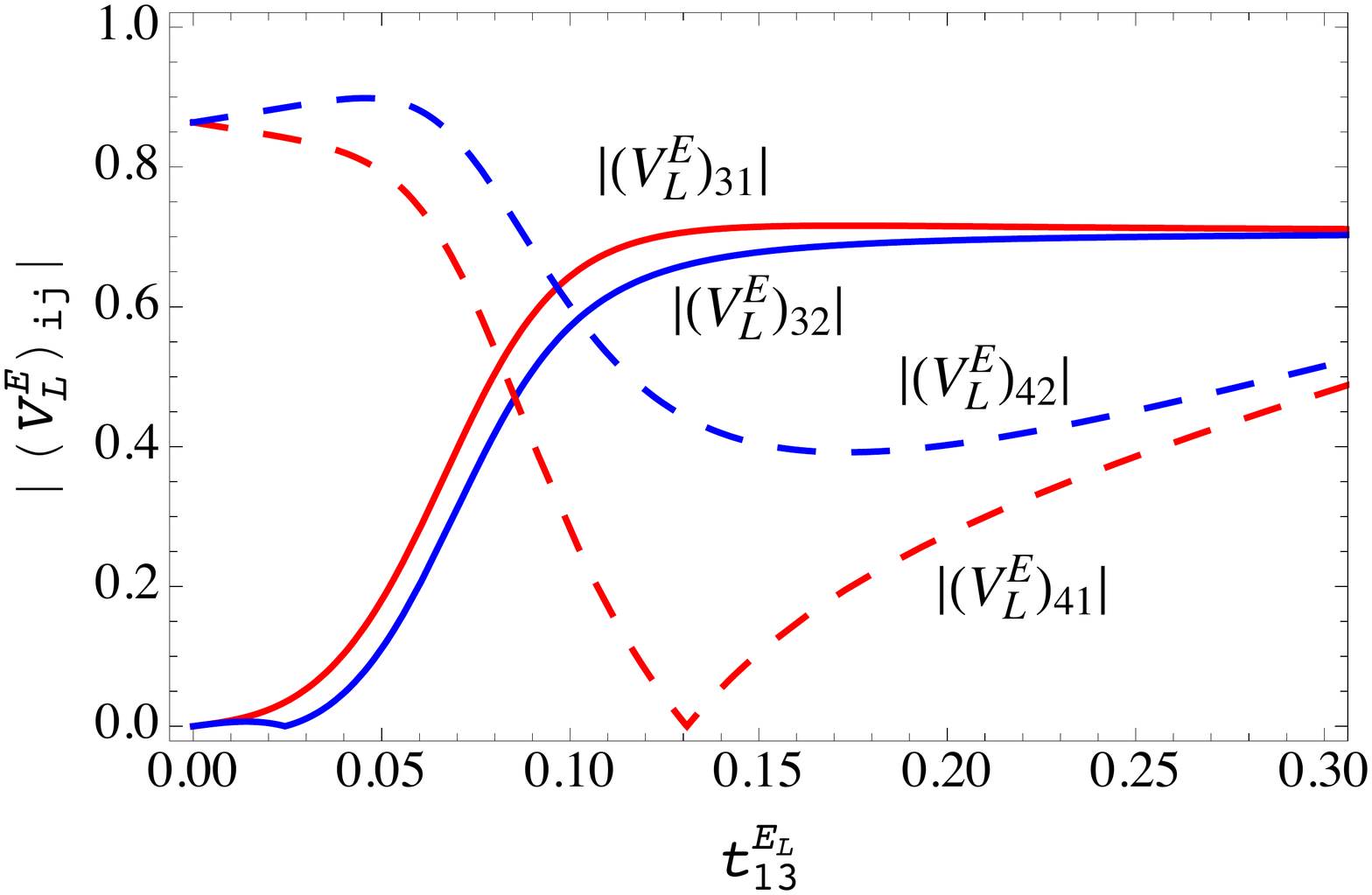}\hspace{-0.2cm}
\includegraphics[scale=0.28]{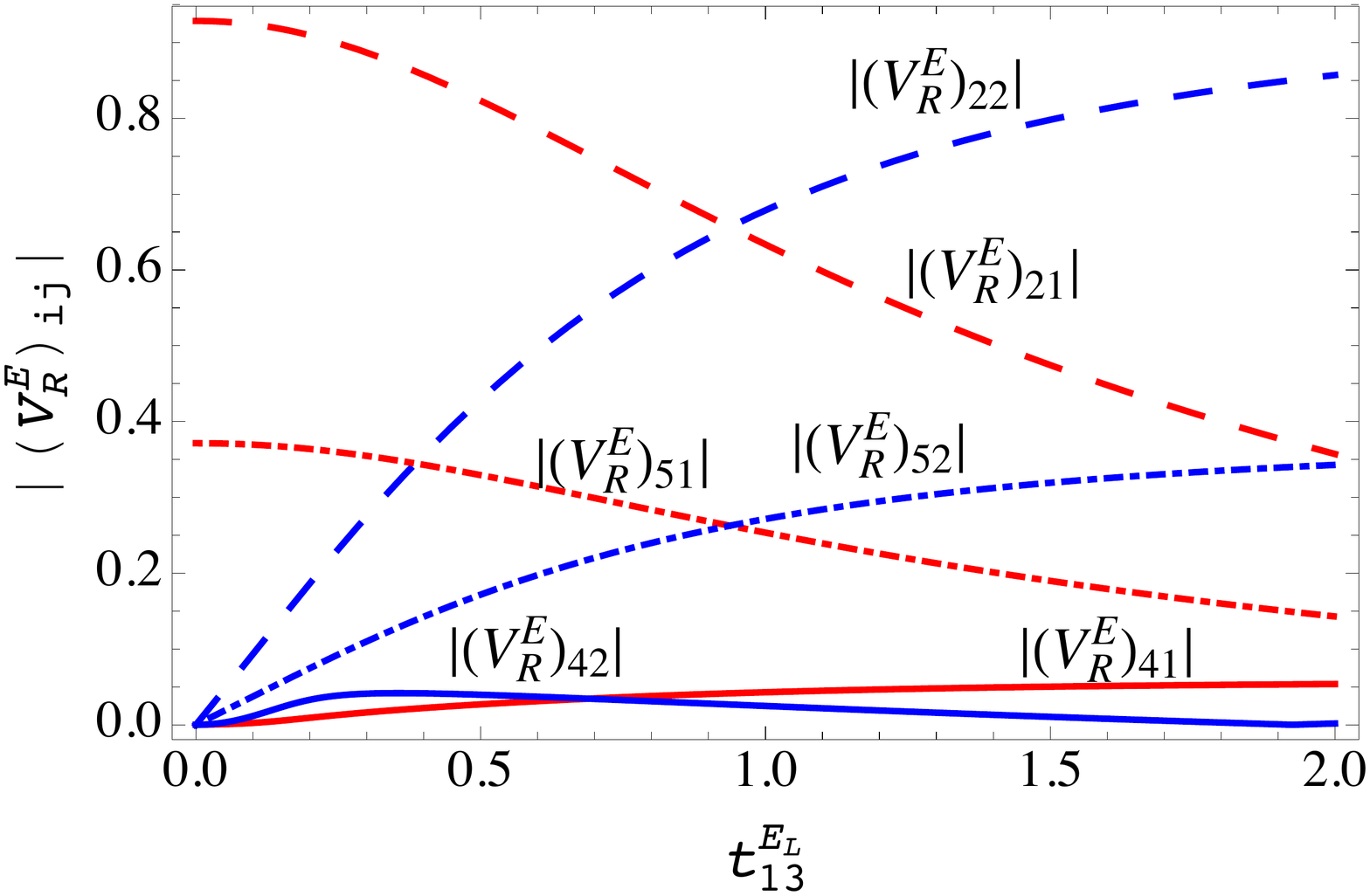}\vspace{-0.5cm}
\caption{ Left-handed ($V_L^{E}$) and right-handed ($V_R^{E}$) biunitary components as function of the mixing tangen $t_{13}^{E_L}$ obtained from equations (\ref{lepton-left-biuni}) and (\ref{lepton-right-biuni}). Each component $ij$ couple to electrons when $j=1$ (red lines) and to muons when $j=2$ (blue lines).}
\label{fig:biunitary}
\end{figure}

In summary, if we fix the parameters as described above, we are left with three free parameters, two mixing angles and one Yukawa ratio $(t_{13}^{E_{L}}, t_{25}^{E_{R}}, r_h)$, which we fit according to the experimental bound in (\ref{lhcb-ratio}). The first aspect to note is that the couplings to electrons have contributions from the biunitary components $\left(V_L^{E}\right)_{a1}$ for $a=3,4,5$ and $\left(V_R^{E}\right)_{a1}$ for $a=2,4,5$, while the muons couple through $\left(V_L^{E}\right)_{a2}$ and $\left(V_R^{E}\right)_{a2}$, as can be verified in tables \ref{tab:NCBLdecay} and \ref{tab:NCBRdecay}. So, the flavour nonuniversality in the model arise from the difference between the $a1$ and $a2$ components of the biunitary matrices, which occur according to equations (\ref{lepton-left-biuni}) and (\ref{lepton-right-biuni}). The plots in figure \ref{fig:biunitary} highlight the difference between electrons and muons components as function of the mixing tangent $t_{13}^{E_{L}}$, where we have fixed the other parameters in an arbitrary form, which only will shift the curves but does not change their fundamental form. We see that for the left-handed leptons in the first plot, the $31$ (red continuous curve) and $32$ (blue continuous curve) components exhibit a small difference, which favoured a universal lepton coupling. The largest lepton universality violation occur due to the $41$ and $42$ components near to $t_{13}^{E_L}=0.13$. The right-handed leptons, on the other hand, exhibit larger violation terms than the left-handed ones, due mainly to the $21$ and $22$ components, as shown in the second plot. The largest differences occur for $t_{13}^{E_L}$ far from 1, which may generate two scenarios: for small and for large $t_{13}^{E_L}$ mixing. However, as we will discuss below, this angle is suppressed as the muon to top quark mass ratio $m_{\mu}/m_t$, thus the scenary with small $t_{13}^{E_L}$ will be favoured.

\begin{figure}
\centering
\includegraphics[scale=0.29]{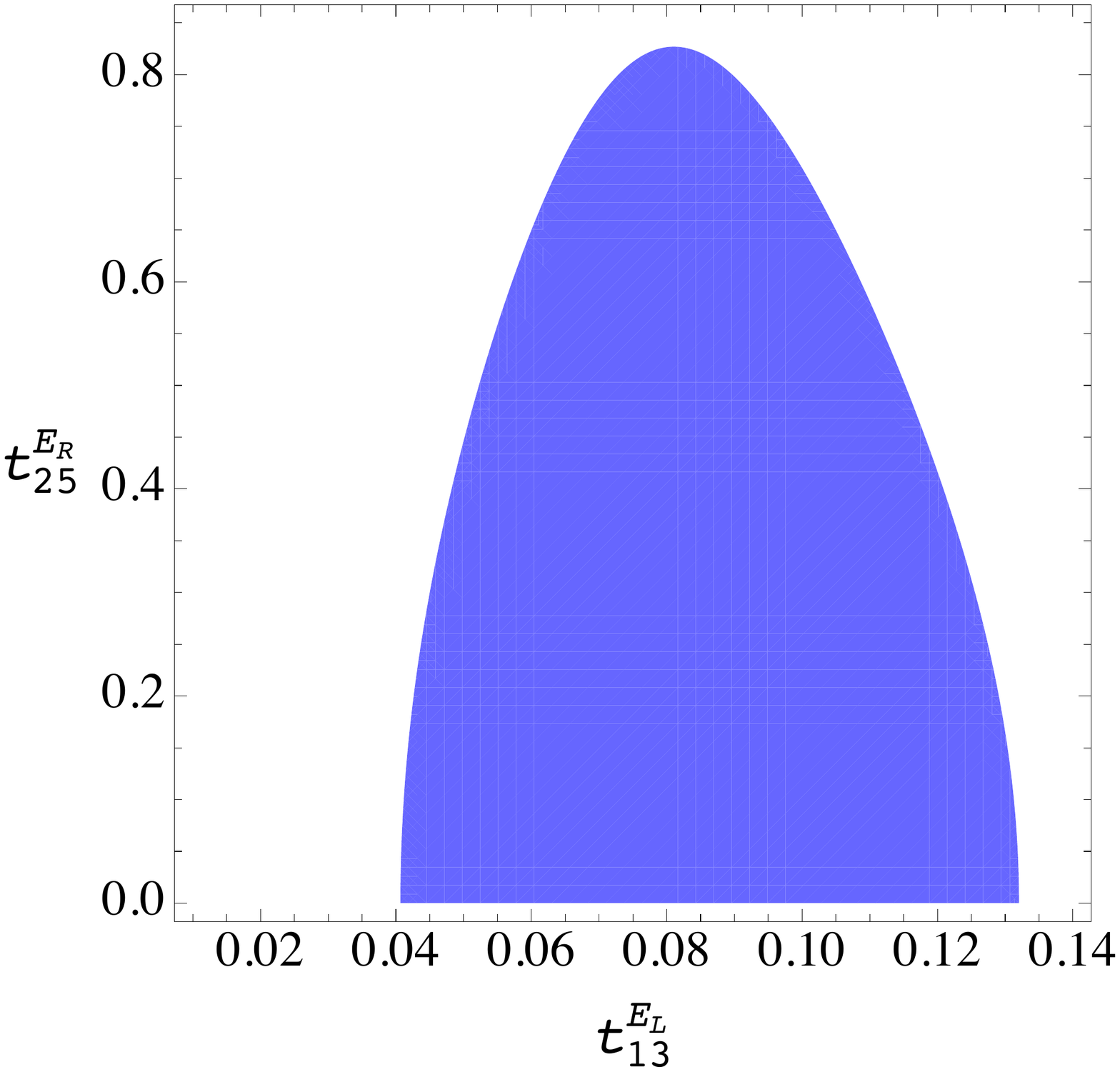}\hspace{-1.7cm}
\includegraphics[scale=0.29]{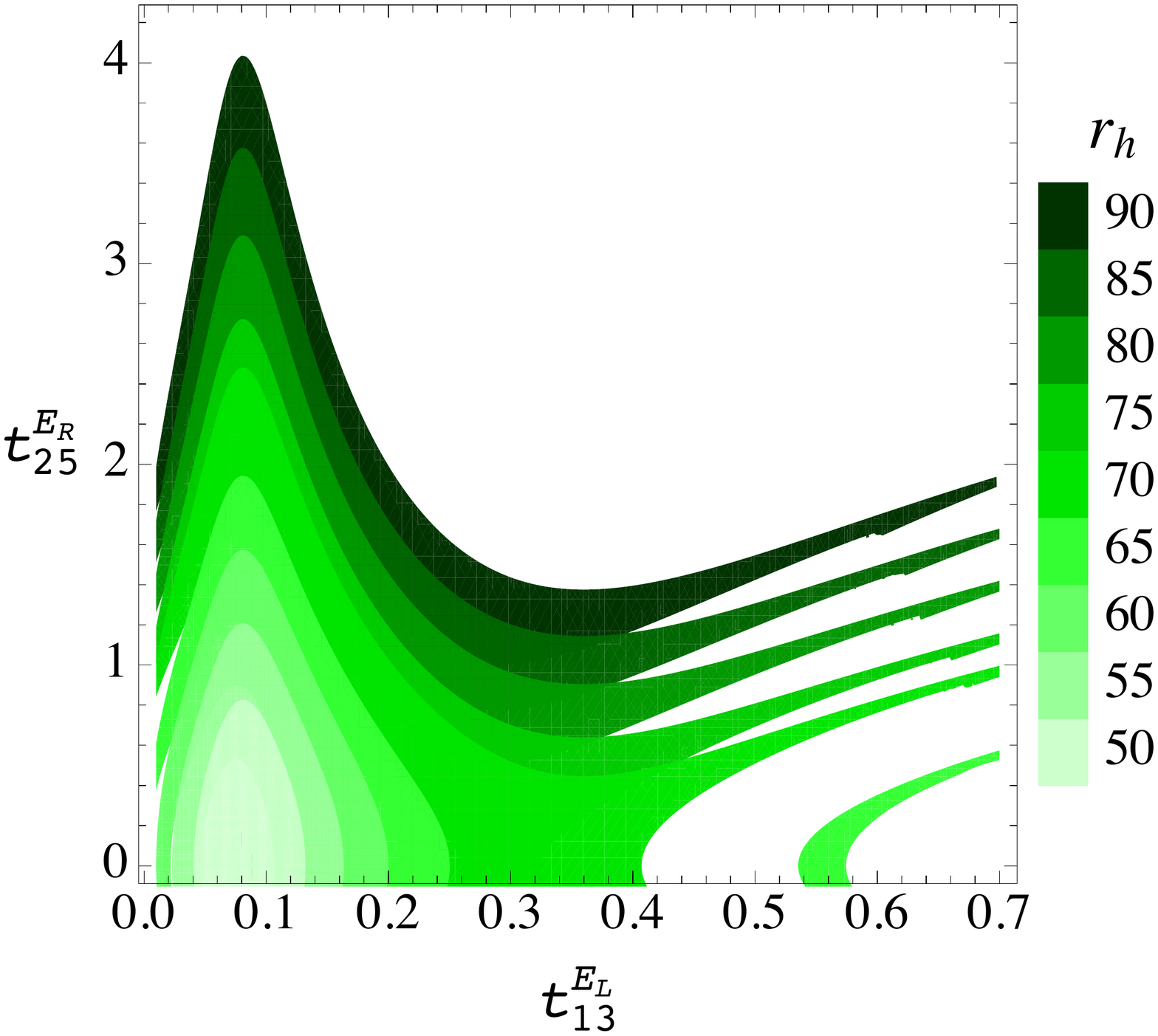}\vspace{-0.5cm}
\caption{Allowed points for the tangent  of the mixing angles $\theta _{13}^{E_{L}}$ and $\theta _{25}^{E_{R}}$ for Yukawa ratio $r_h=50$ (left plot), and for Yukawa ratios spanned from $r_h=50$ to $90$ (right plot) compatible with the experimental limit on $R_K$.}
\label{fig:mixing-region-1}
\end{figure}


Numerically, we found that the reported anomaly can be fitted only for large Yukawa ratios, above $r_h\gtrsim 45$, i.e. the Yukawa couplings that mix the new fermions $\mathcal{J}$ and $\mathcal{E}$ with the ordinary SM fermions must be larger than the couplings among the ordinary up-type quarks in a factor of the order of $4.5\times 10^1$, which corresponds to the order of the absolute values if we assume couplings of the ordinary particles at the order of 1. An important implication to have large Yukawa couplings is the possibility to find a Landau pole in the Yukawa coupling below the Planck scale, which would reduce the perturbative regimen of the model. A deep analysis in this aspect require  a careful study of the renormalization group equations of the theory, which falls outside the scope of this work.

 Regarding the mixing angles, the left plot in figure \ref{fig:mixing-region-1} displays allowed points in the $(t_{13}^{E_{L}}, t_{25}^{E_{R}})$ plane for $r_h = 50$, where a small but non-null mixing angle $\theta _{13}^{E_{L}}$ is require, while $\theta _{25}^{E_{R}}$ can be as large as $42^{0}$, which occur for $\theta _{13}^{E_{L}}\approx 4.6^{0}$. According to (\ref{right-miximg-angles-large}), a $\theta _{25}^{E_{R}}$ mixing angle near $45^0$ (i.e, $t_{25}^{E_R}\sim 1$ ) represents an scenary where all the couplings with the new leptons $\mathcal{E}$ have the same strenght. However, most of the allowed points spread around a small $25$ mixing, where the couplings of the new leptons is larger than their mixing coupling with the ordinary leptons. On the other hand, small $\theta _{13}^{E_{L}}$ mixing is expected according to (\ref{eq:Electron-SM-Rotation-angles}), where the tangent of this angle is proportional to the VEV ratio $v_3/v_1$. Since $v_1$ is proportional to the top quark mass, while $v_3$ is proportional to the muon mass as seen in equations (\ref{top-mass}) and (\ref{lepton-mass-natural}) , then this mixing angle is suppresed by the ratio $m_{\mu}/m_t$. If we increase the Yukawa ratio $r_h$, larger mixing angles can be obtained. The plot in the right of figure \ref{fig:mixing-region-1} shows contourplots for different ratios $r_h$ from 50 to 90. Regarding the other mixing angles, they can be obtained from equations (\ref{leptonmixing-natural}) and (\ref{see-saw-left-lepton-2}) once $\theta _{13}^{E_{L}}$ and $\theta _{25}^{E_{R}}$ are fixed in accordance with the above allowed regions.
 
\begin{figure}
\centering
\includegraphics[scale=0.35]{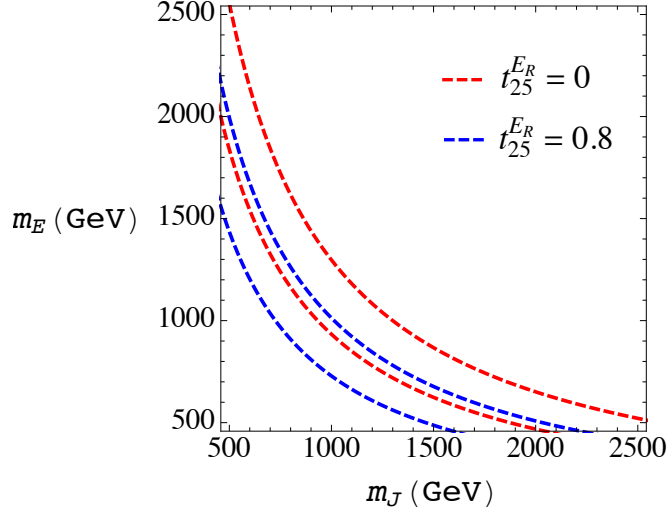}
\vspace{-0.5cm}
\caption{Closed contours in the $(m_J, m_E)$ plane for the extra fermion masses with central value $t_{13}^{E_L}=0.08$ and the two limits $t_{25}^{E_R}=0$ and $0.8$, compatible with the allowed region from figure \ref{fig:mixing-region-1}.} 
\label{fig:mass-region}
\end{figure}

On the other hand, the branching ratio is also very sensitive to the masses of the extra fermions, $m_E$ and $m_J$. To explore this, in figure \ref{fig:mass-region} we display the allowed contours for the heavy quarks and charged leptons compatible with the limits in figure \ref{fig:mixing-region-1} for $r_h=50$. We choose the two limits for the $\theta _{25}^{E_{R}}$ angle, at $0$ and $0.8$, for the central value $\theta _{13}^{E_{L}}=0.08$. We see that large mass values of one fermion, require smaller masses of the other one, which is confined in an energy range attainable by the LHC.  Thus, the anomaly in the meson decay is compatible with new physics at the TeV scale. 

\begin{figure}
\centering
\includegraphics[scale=0.35]{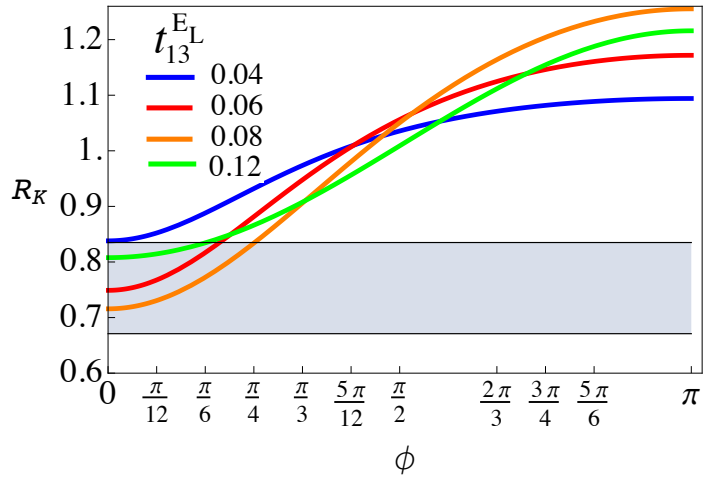}\vspace{-0.5cm}
\caption{Muon to electron branching ratio as function of the complex phase of $\tilde{g}_{L3}^{(23)}$ for $t_{25}^{E_R}=0$ and $t_{13}^{E_L}=0.04, 0.06, 0.08,$ and $0.12$. The shaded area is the reported bound.} 
\label{fig:Rk-phase}
\end{figure}

In tha above discussion, we assume real mixing rotations for the mass eigenstate transformations of the fermions. As a results, all the neutral current couplings in tables \ref{tab:NCBLdecay} and \ref{tab:NCBRdecay} take real values. Now we want to explore the role of possible complex phases in the biunitary transformations. For the lepton couplings, we see in  tables \ref{tab:NCBLdecay} and \ref{tab:NCBRdecay} that the mixing matrices contributes as the squared of their magnitudes $\left|\left(V_L^{E}\right)_{ij}\right|$, so any complex phase associated to this sector does not have any effect in the branching ratios. For the quark couplings $\tilde{g}_{Lm}^{(23)}$, we see that they can be complex in general. In particular, if we neglect the $Z-Z'$ mixing angle, the only contribution to the $b\rightarrow s$ transition is the first term of $\tilde{g}_{L3}^{(23)}$, which may provide a relative complex phase between $\left(V_L^{D}\right)_{21}$ and  $\left(V_L^{D}\right)_{13}$, which we call $\phi $. Thus, in this more general scenario, the new physics of the Wilson coefficients in (\ref{NP-WILSON-a}) and (\ref{NP-WILSON}) will have a global complex term $e^{i\phi }$ coming from the coupling $\tilde{g}_{L3}^{(23)}$. If $\phi = 0$, we reproduce the same physics as shown above. If $\phi=\pi$, we obtain again real coefficients, but with opposite relative signs. For $0<\phi<\pi $, the Wilson coefficients will have new complex contributions. In particular, if we take the same parameters as in figure \ref{fig:mixing-region-1}, we can evaluate the ratio $R_K$ for different values of the complex phase. For example, figure \ref{fig:Rk-phase} shows the branching ratio as a function of the phase for  $r_h=50$ , $t_{25}^{E_R}=0$ and $t_{13}^{E_L}$ between the limits $0.04$ and $0.12$. The shaded band is the allowed region according to the reported anomaly. We first see that there are allowed solutions for small complex phases, obtaining the largest value at $\phi=\pi/4$ when $t_{13}^{E_L}=0.08$. Second, we note that for $\phi = \pi$, the curves lies outside the allowed region. Thus, the sign (or more general, the phase) of the new physics contribution is essential to determine the best scenario to explain the observed anomaly.

\section{Model in the decoupling limit}

The mixing couplings with the extra particles matter $\mathcal{E}^{1,2}$, $\mathcal{J}^{1,2}$ and $\mathcal{T}$ occurs through the fermionic biunitary matrices $\left(V_{L,R}\right)_{i\alpha}$, with $i$ the flavor index for the ordinary matter and $\alpha$ for the new matter. In the above section, we highlighted the importance of the new fermions in the simple scenary with ''natural'' parametrization. As a result, relatively large mixing couplings (strong coupling limit) is required in order to fit the observed anomaly of the $B_s$ decay. If we reduce the mixing couplings to zero, i.e., if the $i\alpha$ components of the mass matrices are ignored, then we obtain the decoupling limit, where only ordinary fermions participate in the decay process. In particular, according to (\ref{eq:Electron-mass-matrix}) and (\ref{eq:Electron-SM-Rotation-angles}), the leptonic 13 left-handed mixing tangent would diverge ($t_{13}^{E_L}\rightarrow \infty$) in this limit, while from (\ref{right-miximg-angles-large}) its 25 right-handed tangent would cancel out ($t_{25}^{E_R}=0$). Figure \ref{fig:decoupling} displays the branching ratio for different $t_{13}^{E_L}$ values and $t_{25}^{E_R}=0$ as function of the Yukawa ratio $r_h$. We observe that for small $t_{13}^{E_L}$ values (below 1),  there are solutions in the shaded region of the reported interval for $R_K$. However, for  $t_{13}^{E_L}\geq 1$, the theoretical values of $R_K$ increases above the allowed region. In the decoupling limit, with large $\theta_{13}$ angles, the branching ratio goes to the SM limit $R_K^{SM}=1$. Thus, the model in this scenario does not account for the reported anomaly. However, we can relax the natural parametrization to more general cases in order to obtain a feasible scenario in the decoupling limit. For that, we first reparametrize the neutral current couplings from tables \ref{tab:NCBLdecay} and \ref{tab:NCBRdecay} in the decoupling limit as:    

\begin{figure}
\centering
\includegraphics[scale=0.3]{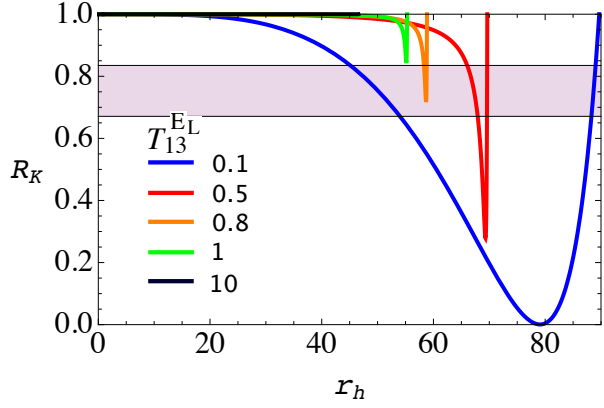}\vspace{-0.5cm}
\caption{Muon to electron branching ratio as function of the Yukawa coupling ratio $r_h$ for $t_{25}^{E_R}=0$ and $t_{13}^{E_L}=0.1, 0.5, 0.8, 1$ and $10$. The shaded area is the ratio experimentally reported in \cite{lhcb}} 
\label{fig:decoupling}
\end{figure}

\begin{eqnarray}
\tilde g_{L2}^{(23)}&=&\frac{2g_X}{3g}\left(V_{L}^{D\dag}\right)_{21}\left(V_{L}^{D}\right)_{13}S_{\theta}, \nonumber \\
\tilde g_{L3}^{(23)}&=&\frac{1}{S_{\theta}}\tilde g_{L2}^{(23)}, \nonumber \\
\tilde g_{L2}^{(aa)}&=&-u_9-\frac{2g_X}{g}\left|\left(V_{L}^{E}\right)_{3a}\right|^2S_{\theta},\nonumber \\
\tilde g_{L3}^{(aa)}&=&u_9S_{\theta}-\frac{2g_X}{g}\left|\left(V_{L}^{E}\right)_{3a}\right|^2, \nonumber \\
\tilde g_{R2}^{(aa)}&=&u_{10}-u_9+\frac{2g_X}{g}\left(-\frac{4}{3}+\left|\left(V_{R}^{E}\right)_{2a}\right|^2\right)S_{\theta}, \nonumber \\
\tilde g_{R3}^{(aa)}&=&\left(u_{9}-u_{10}\right)S_{\theta}+\frac{2g_X}{g}\left(-\frac{4}{3}+\left|\left(V_{R}^{E}\right)_{2a}\right|^2\right),
\end{eqnarray}
with
 
\begin{eqnarray}
 u_9=\frac{1-2S_W^2}{C_W}, \ \ \ \ \ \ \ \ u_{10}=\frac{1}{C_W}.
\end{eqnarray}
By ignoring the $Z-Z'$ mixing angle, the Wilson coefficients for new physics defined by (\ref{NP-WILSON}) become:

\begin{eqnarray}
C_{9}^{NP(a)}&=&\frac{g_X^2\left(36 \text{TeV}\right)^2}{8M_{Z'}^2}K_9^{(a)}, \ \ \ \ \ \ \ \ \
C_{10}^{NP(a)}=\frac{g_X^2\left(36 \text{TeV}\right)^2}{8M_{Z'}^2}K_{10}^{(a)},
\label{NP-WilsonDecoup}
\end{eqnarray}
where the dependency on the flavour is separated in the coefficients

\begin{eqnarray}
K_9^{(a)}&=&\frac{4}{3}\left(V_{L}^{D\dag}\right)_{21}\left(V_{L}^{D}\right)_{13}\left[\frac{4}{3}-\left|\left(V_{R}^{E}\right)_{2a}\right|^2+\left|\left(V_{L}^{E}\right)_{3a}\right|^2\right], \nonumber \\
K_{10}^{(a)}&=&\frac{4}{3}\left(V_{L}^{D\dag}\right)_{21}\left(V_{L}^{D}\right)_{13}\left[\frac{4}{3}-\left|\left(V_{R}^{E}\right)_{2a}\right|^2-\left|\left(V_{L}^{E}\right)_{3a}\right|^2\right].
\label{K-coeff}
\end{eqnarray}
Thus, the theoretical muon to electron branching ratio in (\ref{Rk-2}) become:

\begin{eqnarray}
R_K=\frac{\left|C_9^{SM}+\frac{g_X^2\left(36 \text{TeV}\right)^2}{8M_{Z'}^2}K_9^{(\mu)}\right|^2+\left|C_{10}^{SM}+\frac{g_X^2\left(36 \text{TeV}\right)^2}{8M_{Z'}^2}K_{10}^{(\mu)}\right|^2}{\left|C_9^{SM}+\frac{g_X^2\left(36 \text{TeV}\right)^2}{8M_{Z'}^2}K_9^{(e)}\right|^2+\left|C_{10}^{SM}+\frac{g_X^2\left(36 \text{TeV}\right)^2}{8M_{Z'}^2}K_{10}^{(e)}\right|^2}.
\label{RK-decoupling}
\end{eqnarray}

In order to compare with the experimental data, we define the new physics deviation as:

\begin{eqnarray}
\Delta C_{a}=\sqrt{\left|C_9^{SM}+C_9^{NP(a)}\right|^2+\left|C_{10}^{SM}+C_{10}^{NP(a)}\right|^2}-\sqrt{\left|C_9^{SM}\right|^2+\left|C_{10}^{SM}\right|^2},
\label{NP-deviation}
\end{eqnarray}
so, the ratio (\ref{RK-decoupling}) become:

\begin{figure}
\centering
\includegraphics[scale=0.3]{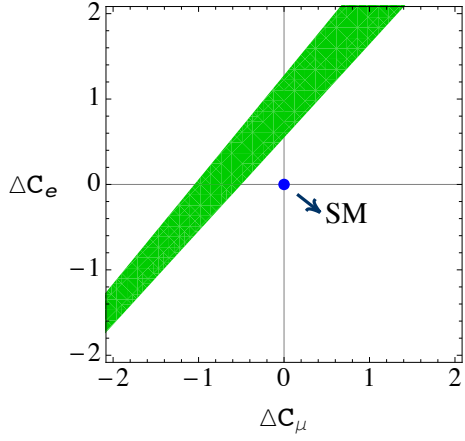}\vspace{-0.5cm}
\caption{Allowed region for the muon and electron new physics deviation defined as equation (\ref{NP-deviation}) compatible with the experimental data. The central blue point is the SM limit} 
\label{fig:Rk-region}
\end{figure}

\begin{eqnarray}
R_K=\left(\frac{\sqrt{\left|C_9^{SM}\right|^2+\left|C_{10}^{SM}\right|^2}+\Delta C_{\mu}}{\sqrt{\left|C_9^{SM}\right|^2+\left|C_{10}^{SM}\right|^2}+\Delta C_{e}}\right)^2.
\end{eqnarray}
Taking into account that $C_9^{SM}\approx -C_{10}^{SM}\approx 4.1$, and the range for $R_K$ in (\ref{lhcb-ratio}), we find in figure \ref{fig:Rk-region} the allowed region for the new physics deviations for muons and electrons, where the SM limit outside the region is shown. We must to compare the above region with the theoretical deviation, determined by the definition (\ref{NP-deviation}) and the parameters from (\ref{NP-WilsonDecoup}). For convenience, we redefine some parameters. First, we define the effective flavour $U(1)_X$ coupling constants as:

\begin{eqnarray}
\left(g_X^{(a)}\right)^2=g_X^2K_9^{(a)}.
\label{efective-coupling}
\end{eqnarray}
Second, we define the two ratios:

\begin{eqnarray}
P_a=\frac{C_{10}^{NP(a)}}{C_{9}^{NP(a)}}, \  \  \  \  \ \ K_{21}=\frac{C_9^{NP(\mu)}}{C_{9}^{NP(e)}}.
\label{Wilson-ratios}
\end{eqnarray}
Thus, the new physics contribution for the ninth electron Wilson coefficient is:

\begin{eqnarray}
C_9^{NP(e)}=\frac{\left(g_X^{(e)}\right)^2\left(36 \text{TeV}\right)^2}{8M_{Z'}^2},
\label{9-wilson}
\end{eqnarray}
while all the remaining coefficients can be parametrized entirely as functions of this as:

\begin{eqnarray}
C_9^{NP(\mu)}=K_{21}C_9^{NP(e)}, \ \ \ \ \ C_{10}^{NP(e)}=P_eC_9^{NP(e)}, \ \ \ \ \ C_{10}^{NP(\mu)}=P_{\mu}K_{21}C_9^{NP(e)}.
\label{Wilson-repara}
\end{eqnarray}
reducing the space of parameters to $(P_e,P_{\mu}, K_{21}, C_9^{NP(e)})$ which we must to fit in order to obtain the allowed deviations according to figure \ref{fig:Rk-region}. Before doing this, we will show that the model predicts a relation between the parameters $P_e$ and $P_{\mu}$. We see from (\ref{NP-WilsonDecoup}) and the definition in (\ref{K-coeff}) that:

\begin{eqnarray}
\frac{1-P_e}{1-P_{\mu}}=K_{21}\frac{\left|\left(V_{L}^{E}\right)_{31}\right|^2}{\left|\left(V_{L}^{E}\right)_{32}\right|^2},
\label{Pe-mu-ratio1}
\end{eqnarray}
where $\left(V_{L}^{E}\right)_{3a}$ are the 31 and 32 components of the lepton left-handed matrix, that in the decoupling limit takes the form:

\begin{eqnarray}
V_L^{E}=
\begin{pmatrix}
V_{SM}^{E} & 0 \\
0 & V_{new}^E
\end{pmatrix},
\end{eqnarray}  
with:

\begin{eqnarray}
V_{SM}^{E}=R(\theta_{23}^{E_L})R(\theta_{13}^{E_L})R(\theta_{12}^{E_L}),
\end{eqnarray}
where each rotation matrix $R(\theta)$ takes the same form as equations (\ref{subrotations-up}) for the quarks, and each angle is defined in (\ref{eq:Electron-SM-Rotation-angles}). In particular, we find for the 31 and 32 components that:

\begin{eqnarray}
\left(V_{L}^{E}\right)_{31}=-s_{12}^{E_L}, \ \ \ \ \ \ \ \left(V_{L}^{E}\right)_{32}=c_{12}^{E_L},
\end{eqnarray}
so that (\ref{Pe-mu-ratio1}) become:

\begin{eqnarray}
\frac{1-P_e}{1-P_{\mu}}=K_{21}\left|t_{12}^{E_L}\right|^2.
\label{Pe-mu-ratio2}
\end{eqnarray}
This condition is equivalent to:

\begin{eqnarray}
\left(C_9^{NP(e)}-C_{10}^{NP(e)}\right)/\left(C_9^{NP(\mu)}-C_{10}^{NP(\mu)}\right)=\left|t_{12}^{E_L}\right|^2.
\label{Pe-mu-ratio3}
\end{eqnarray}

\begin{figure}
\centering
\includegraphics[scale=0.22]{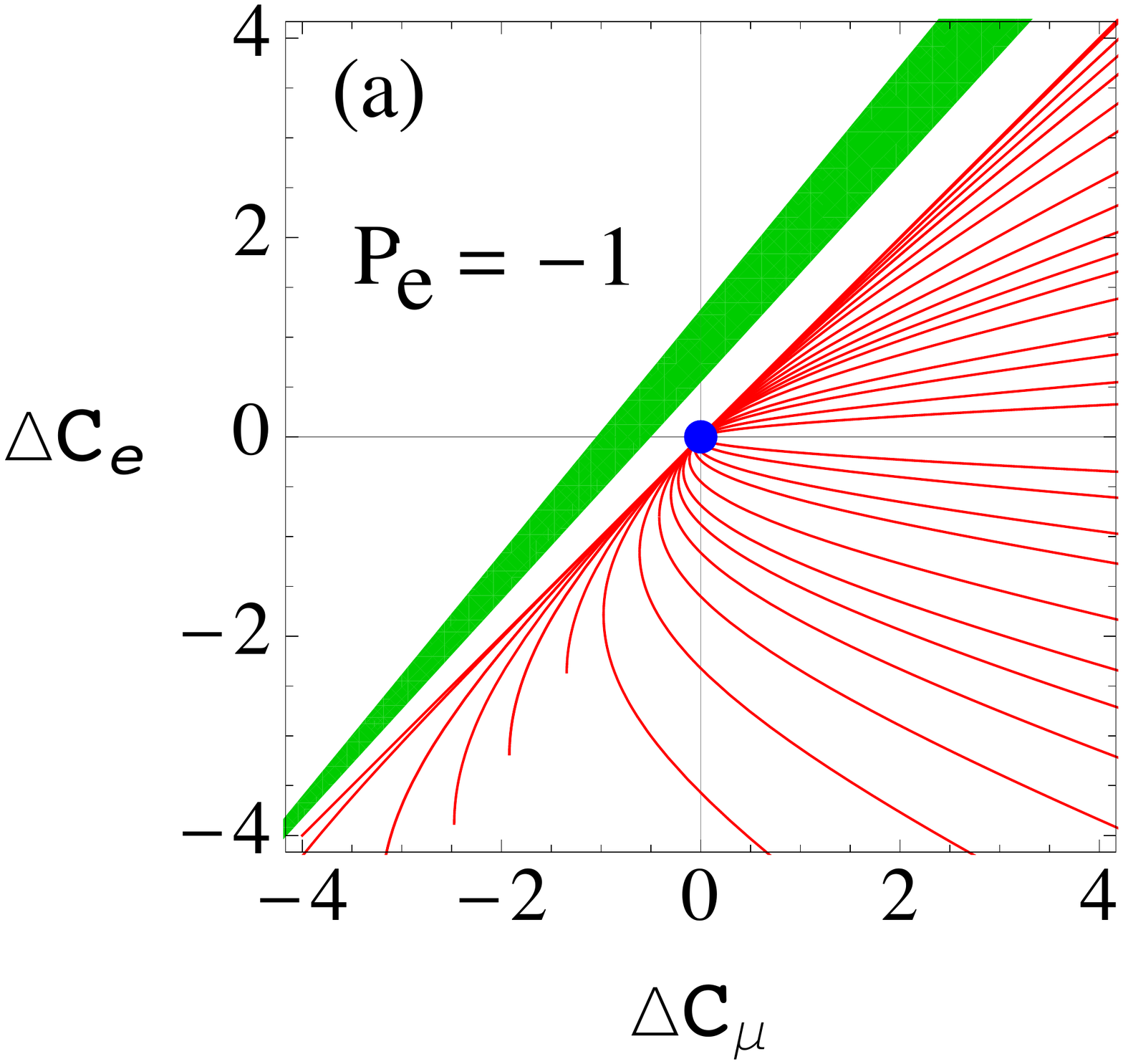}\hspace{-1.7cm}
\includegraphics[scale=0.22]{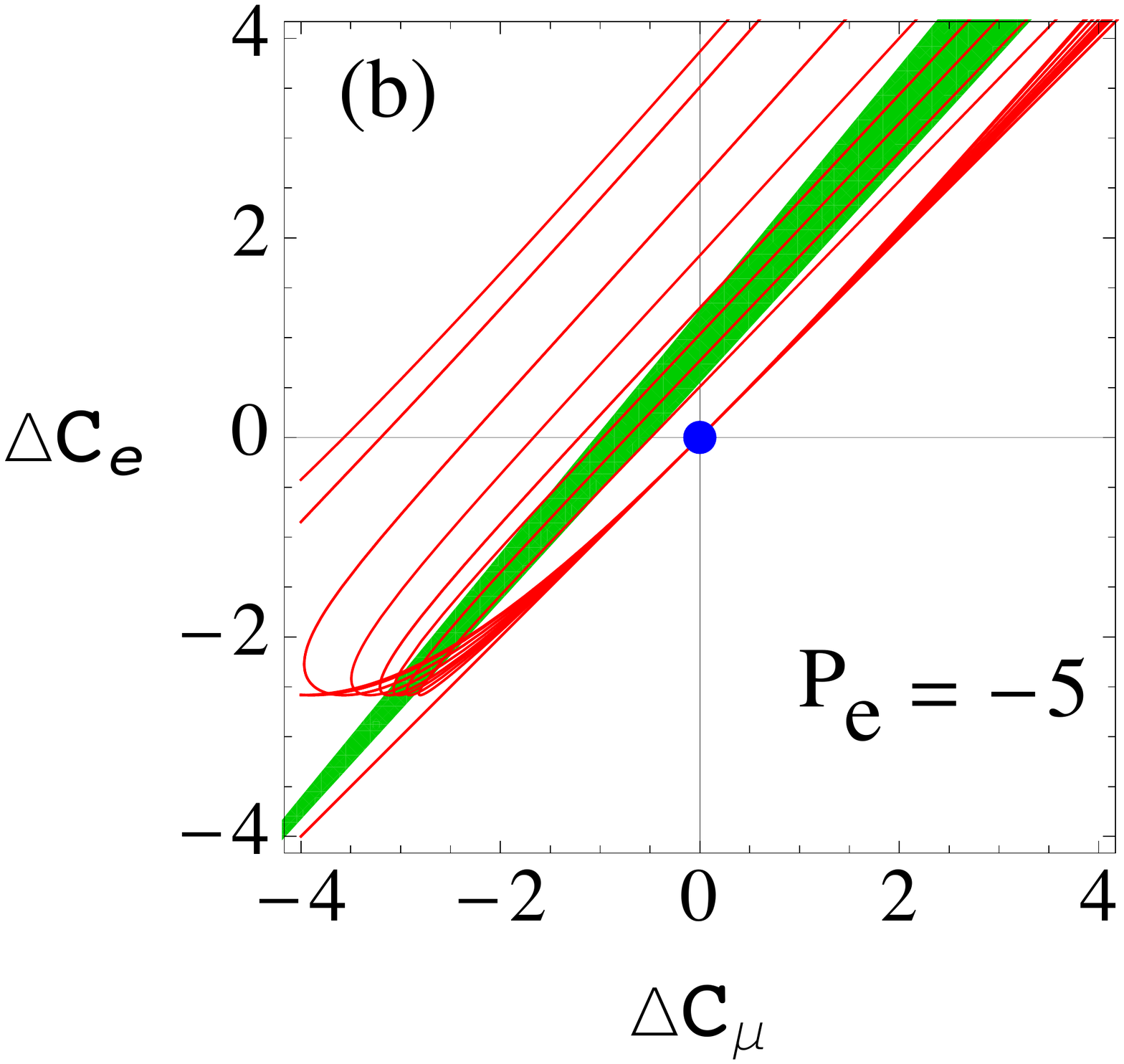}\hspace{-1.7cm}
\includegraphics[scale=0.22]{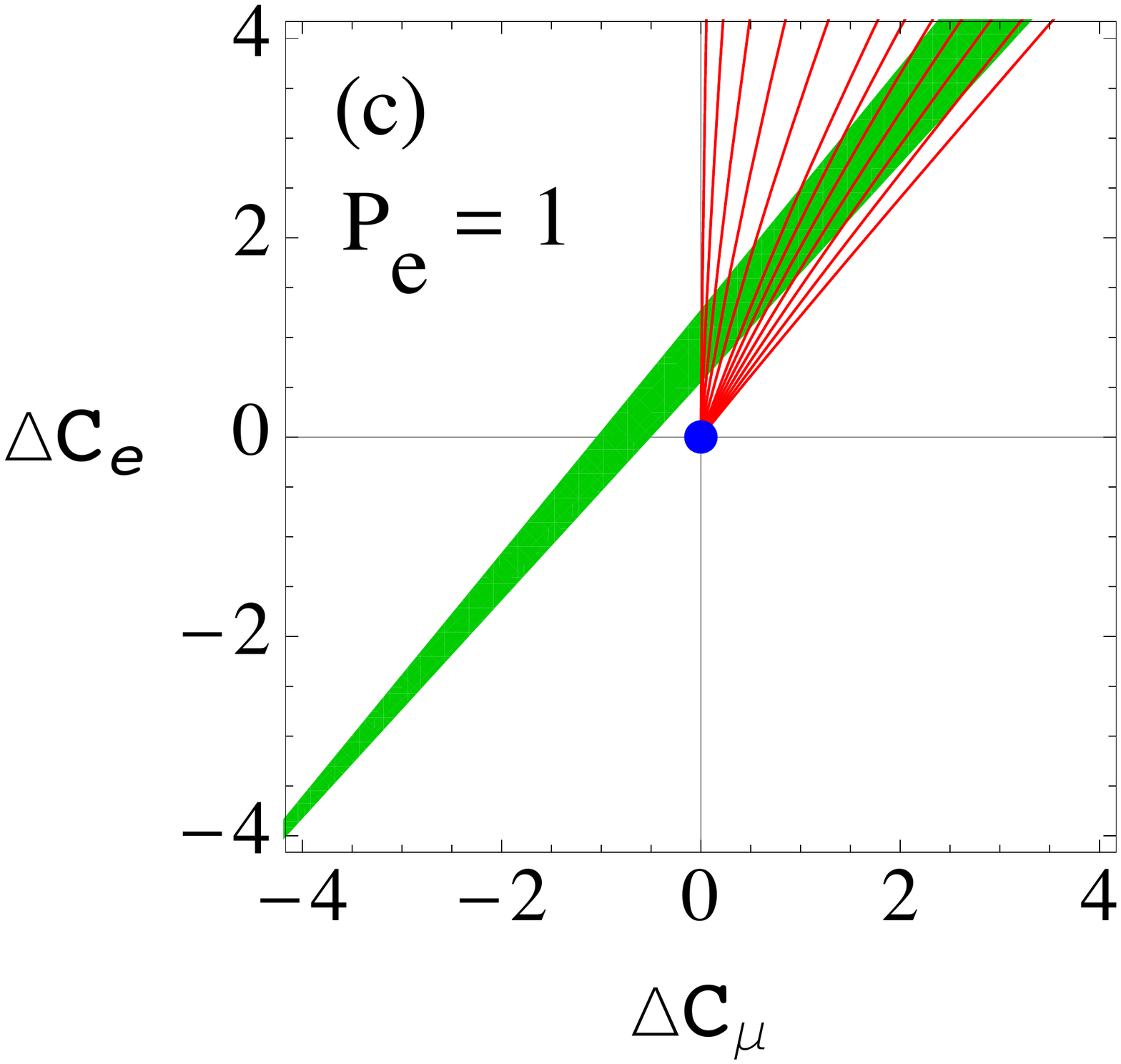}
\vspace{-0.5cm}
\caption{New physics deviations of the Wilson coefficients for $t_{12}^{E_L}=1$ and different values of $P_e$. In (a), there are not solutions through the allowed region for any value of $K_{21}$. In (b) and (c) solutions are found for $1.2\leq K_{21} \leq 5$ and $0 < K_{21} \leq 0.9$, respectively. All the theoretical curves cross the SM limit (blue central point)} 
\label{fig:NP-wilson}
\end{figure}

According to (\ref{leptonmixing-natural}), the limit $t_{12}^{E_L}=1$ is assumed in the natural parametrization. If in addition $P_e=-1$, we obtain for the new physics the same SM relation between the Wilson coefficients: $C_9^{NP(e)}=-C_{10}^{NP(e)}$. However, we did not find any allowed solution on this situation, as shown in graph (a) of figure \ref{fig:NP-wilson}, where the curves are the theoretical predictions for $K_{21}$ ranging from 0 to very large values ($K_{21}\rightarrow \infty$). However, if we deviate from this scenario by choosing other values for $P_e$, we may fit the parameters into the anomaly region in the decoupling limit. For example, the graph (b) in the same figure displays the theoretical solutions for $P_e=-5$ where solutions into the allowed region are found in the interval $K_{21}=[1.2,5]$. From the plot, we can estimate the bound $\Delta C_e\geq -2.6$ for the electron, while for the muon we obtain the allowed interval $-3.2 \leq \Delta C_{\mu} \leq -2.9$ when the former obtains its minimum value. Graph (c) shows the solutions for $P_e=1$ for the interval $0 < K_{21} < 0.9$. Since $K_{21}$ and $t_{12}^{E_L}$ are not zero, according to (\ref{Pe-mu-ratio3}), this case also implies that $P_{\mu} = 1$. Thus, we found scenarios where $C_9^{NP(a)}=C_{10}^{NP(a)}$ for both $a=e$ and $\mu $. In the limit $K_{21}\rightarrow 0$, corrections for the muon $\Delta C_{\mu}$ does not exists, while for electron the allowed range according to graph (c) is $0.5 \leq \Delta C_e \leq 1.3$.

\begin{figure}
\centering
\includegraphics[scale=0.22]{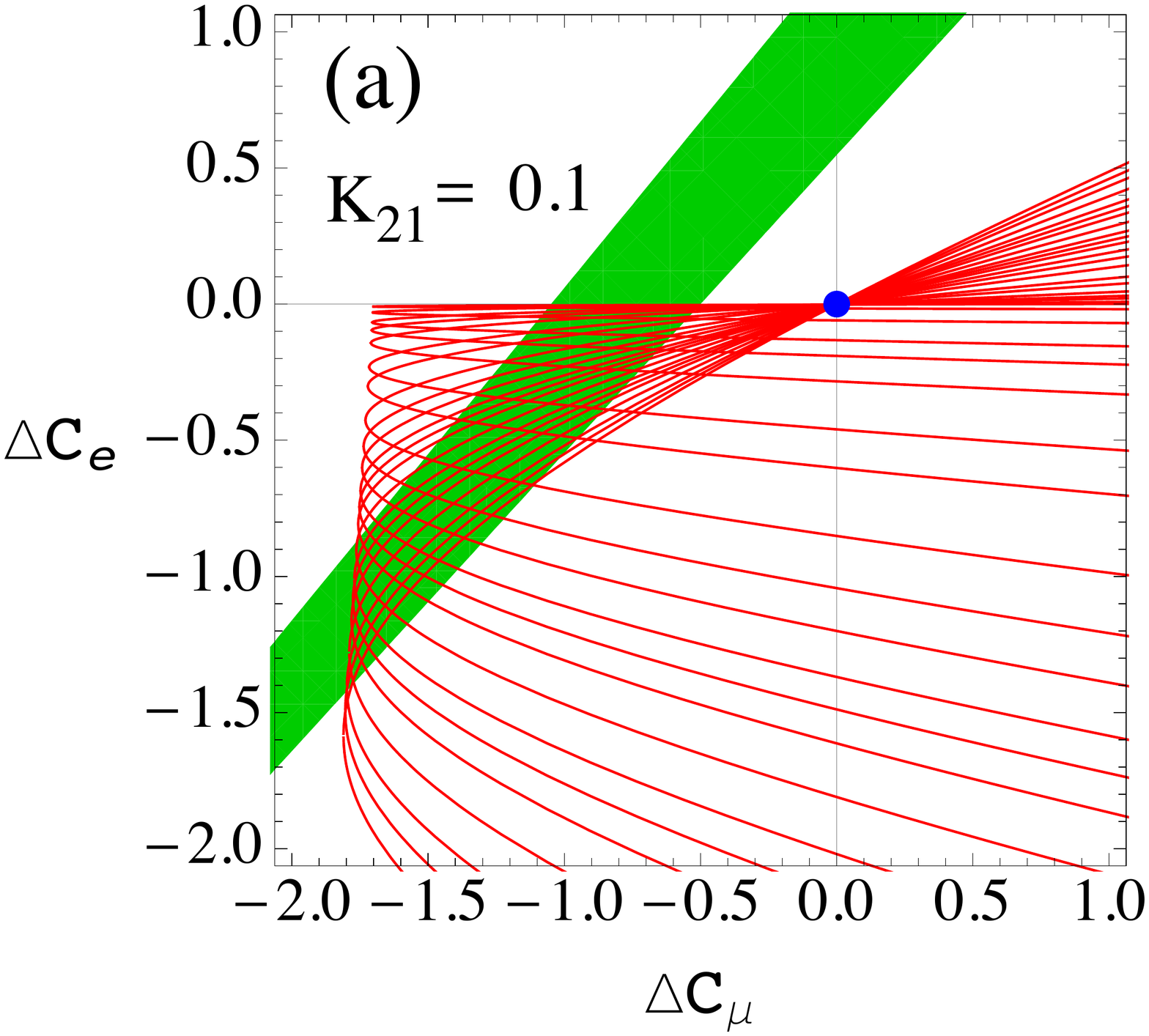}\hspace{-1.6cm}
\includegraphics[scale=0.22]{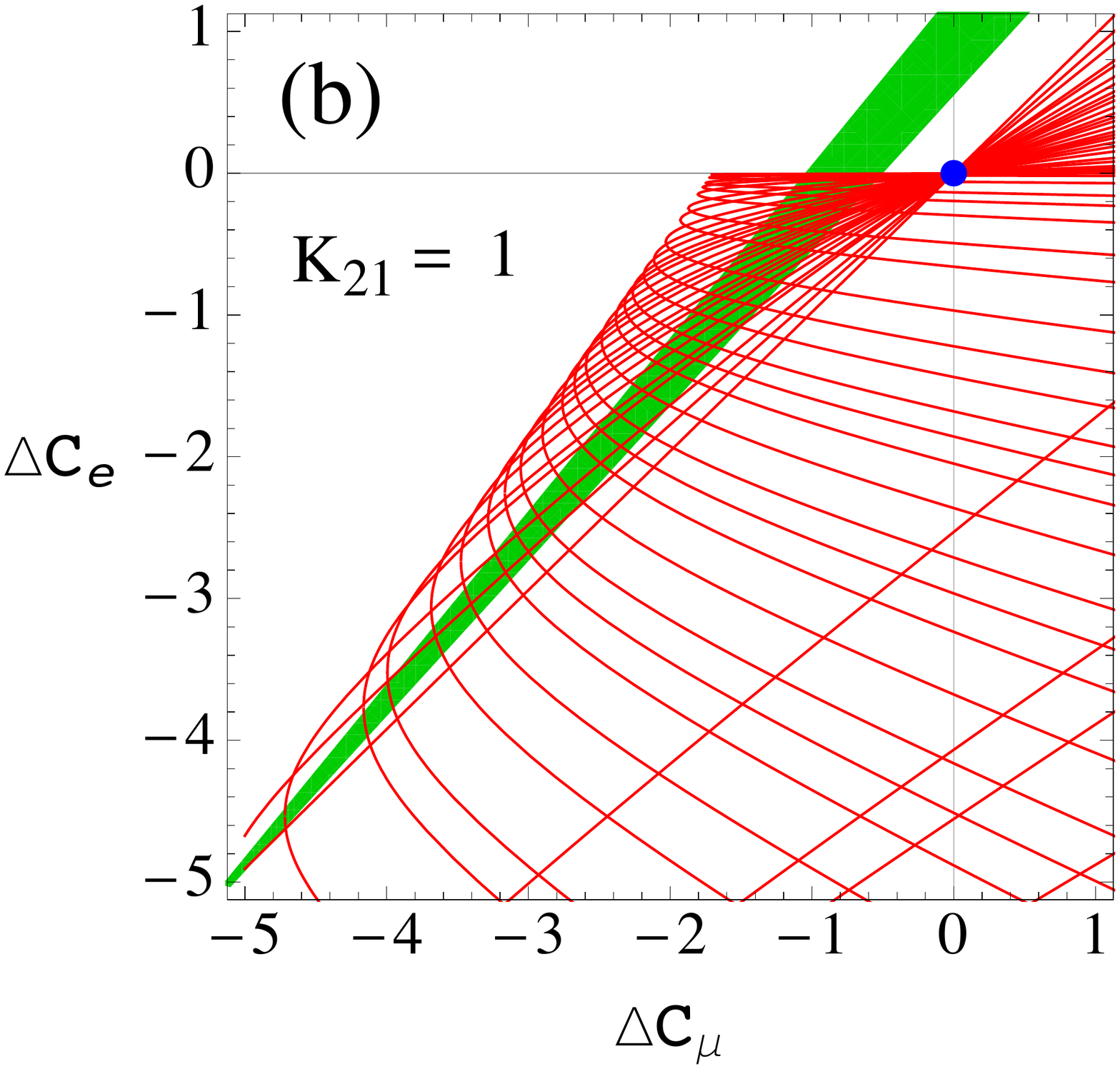}\hspace{-1.6cm}
\includegraphics[scale=0.22]{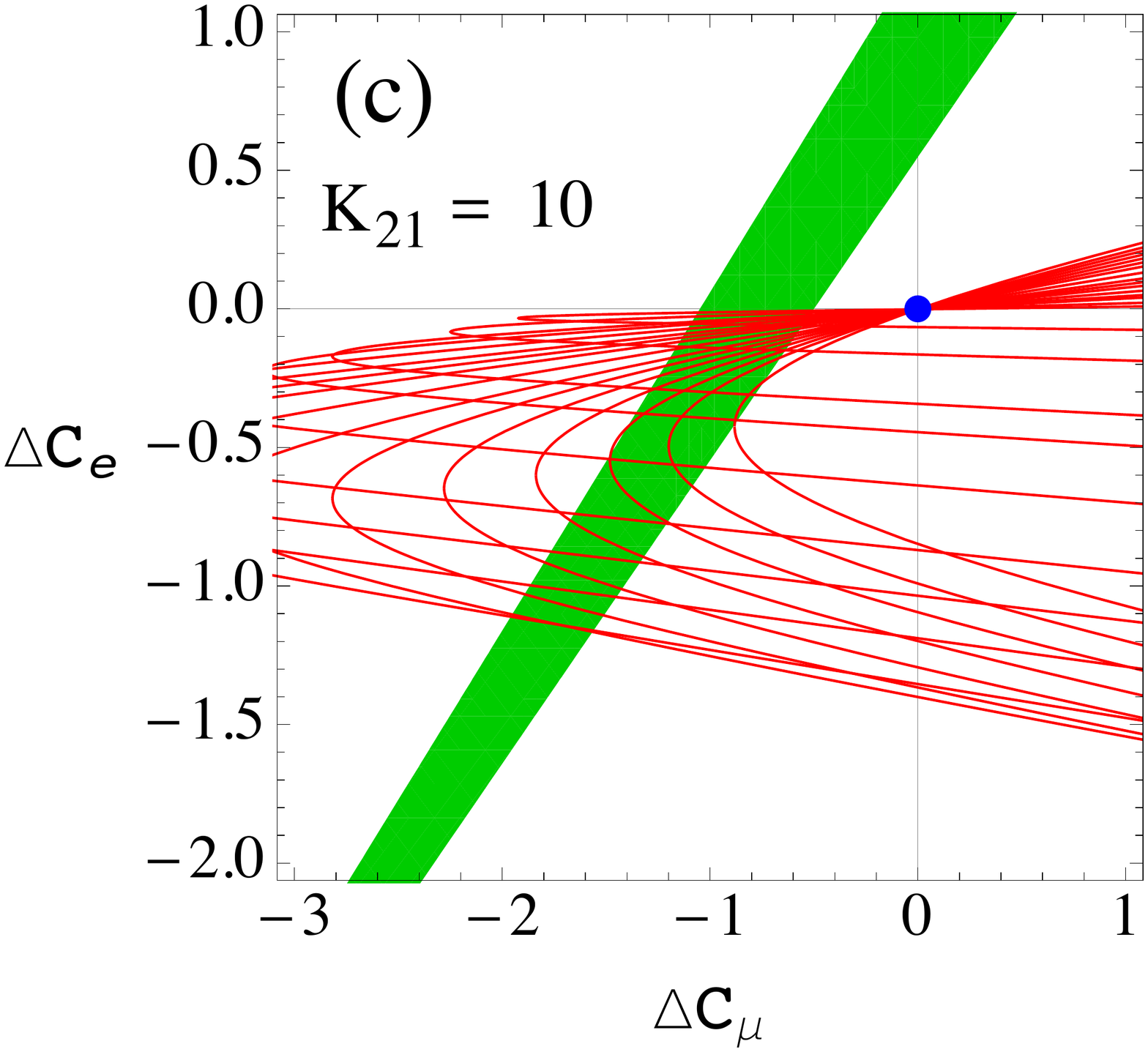}
\vspace{-0.5cm}
\caption{New physics deviations of the Wilson coefficients for $P_{e}=-1$ and (a) $K_{21} = 0.1$, (b) 1 and (c) 10. The curves are for different ranges of $t_{12}^{E_L}$. } 
\label{fig:NP-wilson-2}
\end{figure}

We also may explore scenarios with $t_{12}^{E_L}\neq1$. In particular, the case with $P_e=-1$ can reproduce the reported data by properly fitting the other parameters, as shown in figure \ref{fig:NP-wilson-2}. In graph (a), we obtain solutions for the small ratio $K_{21}=0.1$, and in the range $0 \leq t_{12}^{E_L}\leq 0.72$. Above this limit, the curves falls outside the allowed region, and $\Delta C_{e}=0$ in the limit $ t_{12}^{E_L}=0$. We also see that the curves exhibits the bound $\Delta C_{\mu} \geq -1.8$. In the case with $K_{21}=1$, graph (b) shows a larger range for the deviations, while allowed values extends to the bound $ t_{12}^{E_L} < 1$. For the large value $K_{21}=10$, the curves are shrunk again, as shown in graph (c), where $0 \leq t_{12}^{E_L}\leq 0.51$.

\begin{figure}
\centering
\includegraphics[scale=0.25]{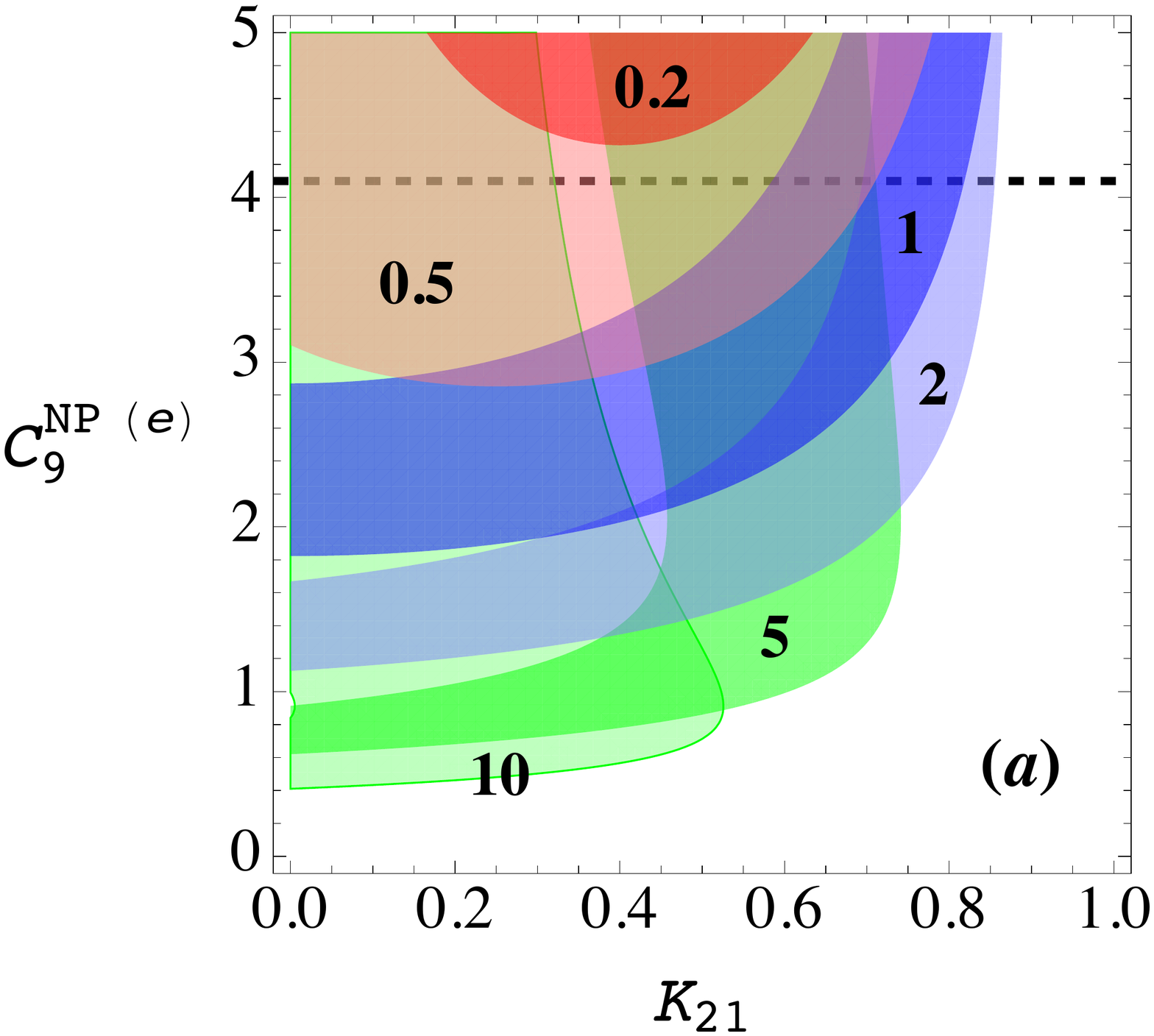} \hspace{2cm}
\includegraphics[scale=0.23]{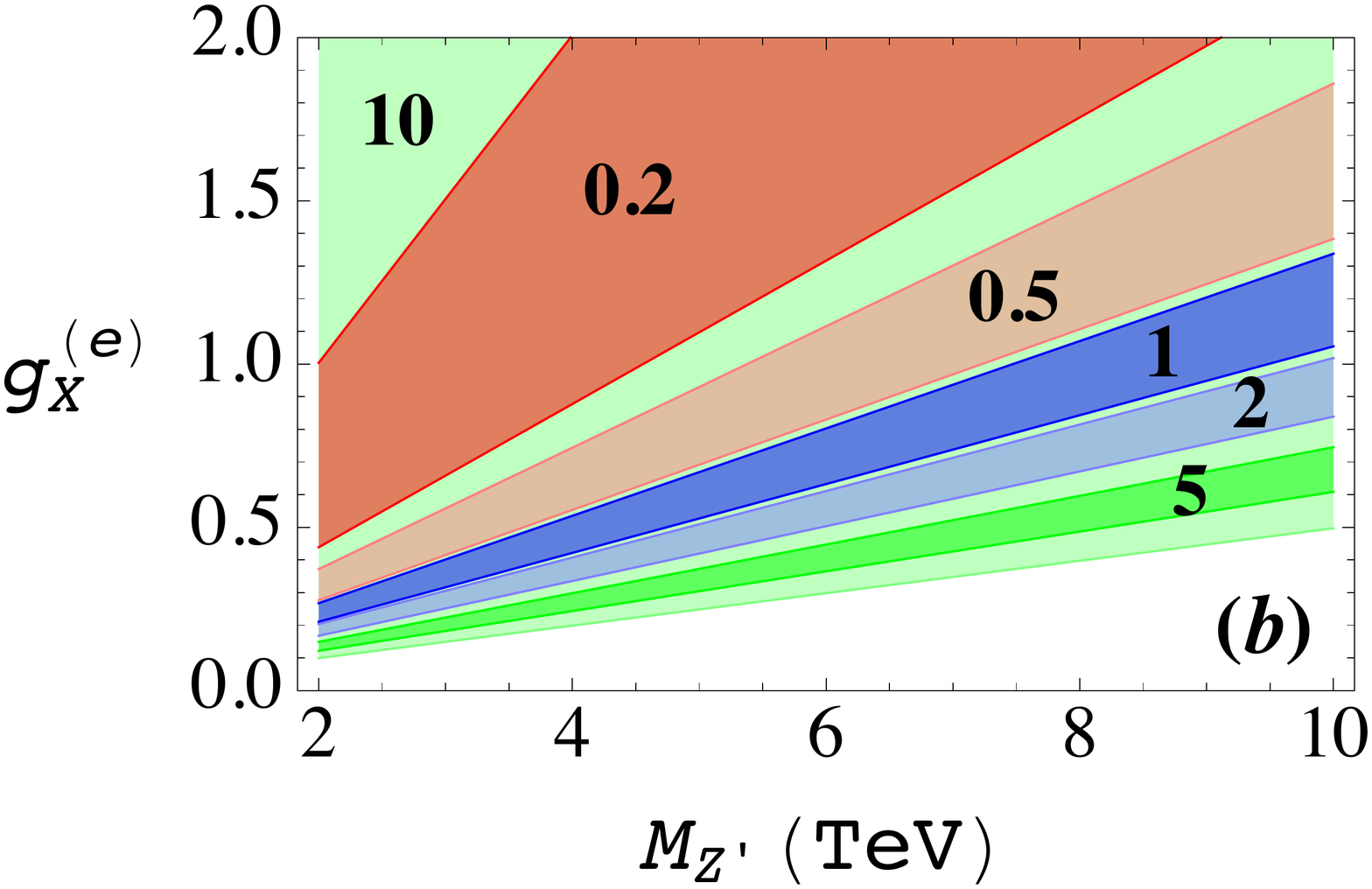} 
\includegraphics[scale=0.23]{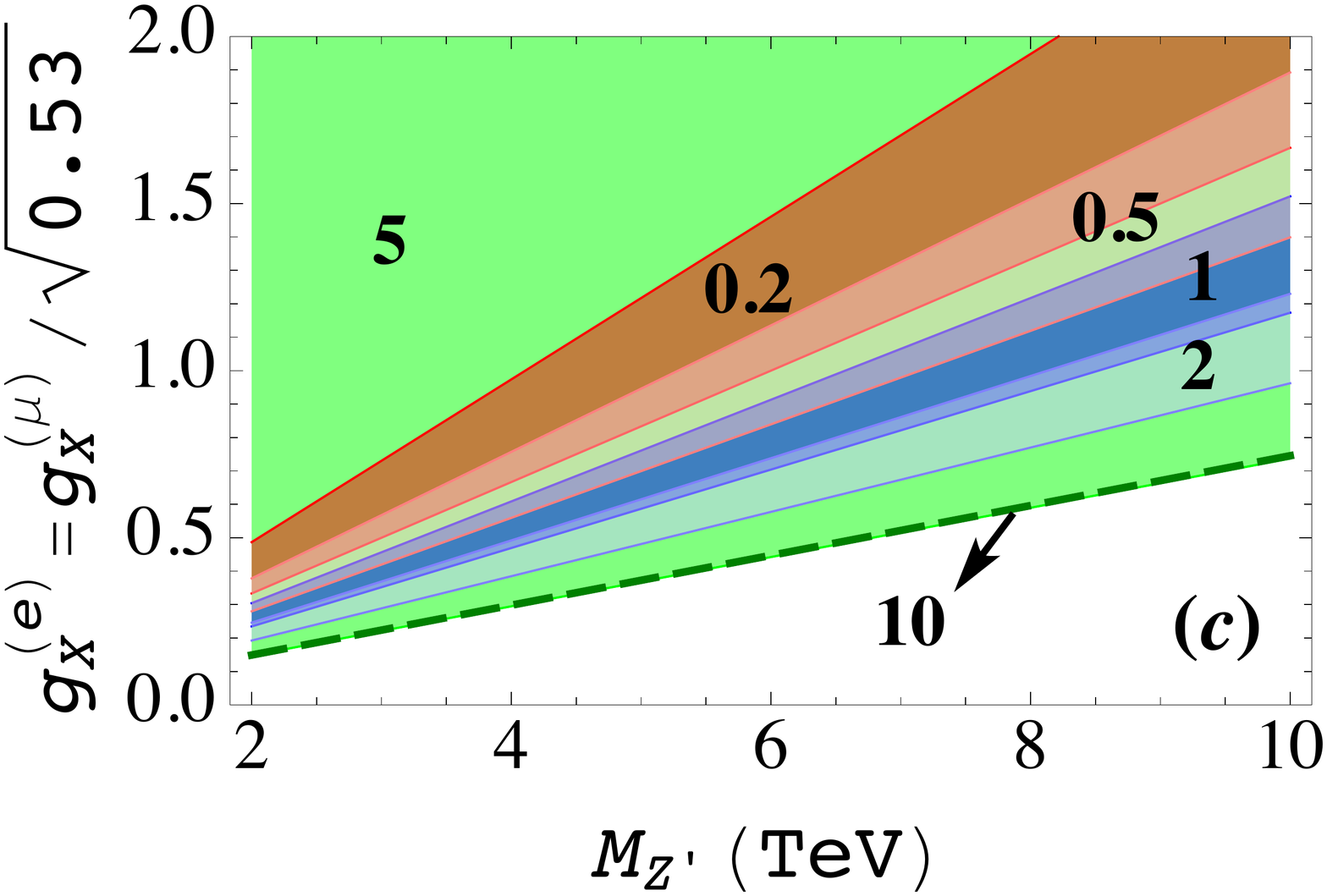}
\vspace{-0.5cm}
\caption{Figure (a) shows allowed regions in the $(C_9^{NP(e)},K_{21})$ plane for $P_{e}=0.2, 0.5, 1, 2, 5$ and $10$. For reference, the dashed horizontal line is the SM limit $C_9^{SM}=4.1$. The plot (b) is the effective coupling as function of the $Z'$ boson mass according to regions in (a) for $K_{21}=0$. Figure (c) is for $K_{21}=0.53$.} 
\label{fig:NP-wilson-3}
\end{figure}

On the other hand, the ratio $K_{21}$ also represents the relative coupling of $e$ and $\mu $ to the $Z'$ boson. Taking into account the equations (\ref{NP-WilsonDecoup}) and the definition (\ref{efective-coupling}), we obtain that:

\begin{eqnarray}
K_{21}=\frac{C_9^{NP(\mu)}}{C_{9}^{NP(e)}}=\frac{\left(g_X^{(\mu )}\right)^2}{\left(g_X^{(e)}\right)^2},
\label{eq:effective-coup-ratio}
\end{eqnarray}
while the Wilson coefficient $C_9^{NP(e)}$ in equation (\ref{9-wilson}) provides a relation between the effective  electron coupling constant $g_{X}^{(e)}$ and the $Z'$ mass. For example, the plot (a) in figure \ref{fig:NP-wilson-3} shows the allowed regions of the electron Wilson coefficient for new physics as function of the ratio $K_{21}=C_9^{NP(\mu)}/C_9^{NP(e)}$, with $t_{12}^{E_L}=1$ and different values of $P_e$: $0.2, 0.5, 1, 2, 5$ and $10$. The dashed horizontal line is the SM limit $C_9^{SM}=4.1$, where we can see that corrections can be smaller, at the same order or, eventually larger than the SM prediction. We see that $K_{21} < 1$, which means that solutions in this scenario are found if electrons couple stronger to the $Z'$ boson than muons. Second, if $C_9^{NP(e)}$ increases, then $P_e$ decreases in accordance with the definition $P_e=C_{10}^{NP(e)}/C_9^{NP(e)}$. So, we see in the plot that the lowest bounds are large for small values of $P_e$. Taking into account these bounds, the plot (b) displays the allowed region for the effective electron coupling $g_{X}^{(e)}$ and the $Z'$ mass for a \textit{muon-phobic} scenario with $K_{21}=0$ ($g_X^{(\mu )}=0$). The plot (c) shows the regions for $K_{21}=0.53$, just at the upper limit of $P_e=10$ as observed in plot (a), and described by the green dashed line in (c). The conversion to the muon coupling is obtained by doing $g^{(e)}_X \times \sqrt{0.53}$, according to (\ref{eq:effective-coup-ratio}).  In general, we see that large ratios $P_e$ favour regions including small gauge couplings constants, which increase as the $Z'$ boson become heavier.

\section{Conclusions}
\label{sect:Conclusions}

Observational facts as the fermion mass hierarchies, mixing schemes, oscillation of neutrinos and experimental anomalies as the $B$ meson decay may be manifestations of new physics beyond the SM. Motivated initially by the fermion mass hierarchy problem, we propose a non-universal U(1)' extension with three Higgs doublets that may reproduce masses and mixing schemes for quarks, charged and neutral leptons. In addition to new charged and neutral Higgs particles, the model introduces other particles from the following conditions:

\begin{enumerate}
\item Due to the new abelian gauge symmetry, a second neutral gauge boson $Z'$ is naturally introduced.

 \item In order to break the $U(1)'$ symmetry and provide mass to the $Z'$ boson, a new Higgs singlet with large VEV is added.
 
  \item Also, the new $Z'$ gauge boson induces quiral anomalies, which may spoil the renormalization of the model. In order to restore the cancellation of these anomalies, we must assign suitable U(1)' charges to the fermions. This assignation is done to obtain flavour non-universal interactions for quarks and leptons, which requires extra quarks and charged leptons.
\end{enumerate} 

The model exhibits lepton universality violation that may explain the B meson decay anomaly into electron and muon pairs reported by the LHCb collaboration. This observable may test the new couplings of the model, in particular, the anomaly is highly sensitive to the new quark and lepton content of the model through their couplings with the Higgs sector. They participate in the meson decay indirectly through their mixing couplings with the ordinary quarks $b$ and $s$, and the charged leptons $e$ and $\mu $. Since these mixings occur in a non-universal form, then the anomaly can be explained and fitted for new physics at the TeV scale, attainable to be proved in the LHC. 

Although we choose an specific scheme to parameterize the mass matrices for fermions and the mixing angles, they are suppressed or enhanced by ratios of VEVs which we preserve in the natural scheme. Specifically, the VEV of the first Higgs doublet determine the scale of the top quark, i.e., $v_1/\sqrt{2}\sim 173$ GeV. The second VEV gives masses to the quark $b$ and the lepton $\tau $ at $v_2/\sqrt{2} \sim 3$ GeV. Finally, the third VEV is of the order of the quark $s$ and lepton $\mu $ mass, at $v_3/\sqrt{2}\sim 0.1 $ GeV. Thus, we expect mixing angles with values of the order of the ratios of the phenomenological fermions measured experimentally independent of the choosen scheme to address the Yukawa free parameters.

\appendix

\section{Block Diagonalization}\label{app:block}

Let us take a generic matrix with arbitrary dimension of the form:
\begin{eqnarray}
{M}^2=
\begin{pmatrix}
A & C \\
C^T & D
\end{pmatrix},
\label{block-matrix}
\end{eqnarray}
with $A, D$ and $C$ sub-matrices whose elements obey the hierarchy 
\begin{eqnarray}
A \ll C \ll D.
\label{block-hierarchy}
\end{eqnarray} 
The matrix (\ref{block-matrix}), as shown in reference \cite{grimus}, can be block diagonalized approximately by a unitary rotation of the form:
\begin{eqnarray}
V=
\begin{pmatrix}
I & F \\
-F^T & I
\end{pmatrix},
\label{approx-rotation}
\end{eqnarray}
where $I$ is an identity matrix, and $F$ a small sub-rotation with $F\ll 1$. Keeping only up to linear terms on $F$, the rotation gives:
\begin{eqnarray}
V^T M^2 V=
\begin{pmatrix}
A-CF^T-FC^T & C+AF-FD \\
C^T+F^TA-DF^T & D+C^TF+F^TC
\end{pmatrix},
\label{rotation}
\end{eqnarray}
which, by definition, must lead us to a diagonal block form
\begin{eqnarray}
m^2=
\begin{pmatrix}
a & 0 \\
0 & d
\end{pmatrix},
\label{diagonal-block}
\end{eqnarray}
with $a$ and $d$ non-diagonal matrices, and $0$ the null matrix. By matching the upper right non-diagonal block in (\ref{rotation}) and (\ref{diagonal-block}), we obtain that $C+AF-FD=0$. Taking into account the hierarchy in (\ref{block-hierarchy}), we may neglect the term with $A$, finding the following approximate solution:
\begin{eqnarray}
F\approx CD^{-1}.
\label{F-subrotation}
\end{eqnarray}

On the other hand, if we match the diagonal blocks in (\ref{rotation}) and (\ref{diagonal-block}), and using the solution (\ref{F-subrotation}), we can obtain the form of the submatrices $a$ and $b$ in terms of the original blocks $A$, $C$ and $D$. We obtain at dominant order that:
\begin{eqnarray}
a &\approx & A-CD^{-1}C^T \nonumber \\
b &\approx & D.
\end{eqnarray}
The above matrices can be diagonalized independently.  

\section{Parametrization of the biunitary matrix transformations}\label{app:biunitary}

In this appendix we obtain the parameters of the biunitary transformations that rotate the flavour fermion basis into mass basis.

\subsection*{Up sector}

From the Yukawa Lagrangian (\ref{eq:Up-Lagrangian}), we obtain the following mass matrix for the up-type quark sector:

\begin{equation}
\mathbb{M}_{U} = \frac{1}{\sqrt{2}}
\begin{pmatrix}
h_{3 u}^{1 1}v_{3}	&	h_{2 u}^{1 2}v_{2}	&h_{3 u}^{1 3}v_{3}	&h_{2 \mathcal{T}}^{1}v_{2}	\\
0	&	h_{1 u}^{2 2}v_{1}	&	0	&h_{1 \mathcal{T}}^{2}v_{1}	\\
h_{1 u}^{3 1}v_{1}	&	0	&	h_{1 u}^{3 3}v_{1}	&	0	\\
0	&	g_{\chi u}^{2}v_{\chi}	&	0	&	g_{\chi \mathcal{T}}v_{\chi}
\end{pmatrix},
\label{app:up-matrices}
\end{equation}
which diagonalizes through the biunitary matrices $V_{L(R)}^U$. In particular, as shown in reference \cite{1s3d-nonuniversal}, the left-handed matrix can be expressed as the product of two mixing matrices of the form:

\begin{eqnarray}
V_L^{U}=
\begin{pmatrix}
1 & \Theta _L^{U\dag} \\
- \Theta _L^{U} & 1
\end{pmatrix}
\begin{pmatrix}
V_{\text{SM}}^{U} & 0 \\
0 & V_{\text{new}}^{U}
\end{pmatrix},
\end{eqnarray}
where $ \Theta _L^{U} $ is a see-saw matrix that block-diagonalize the mass matrix into one mass matrix of the ordinary SM quarks and a heavy matrix that mixes the new quarks, while $V_{\text{SM}}^{U} $ and $V_{\text{new}}^{U}$ diagonalize each of these matrices. For simplicity, we assume diagonal exotic matrices, so that $V_{\text{new}}^{f}=1$.  The see-saw matrix is:

\begin{equation}
\Theta^{U\dagger}_{L} = 
\begin{pmatrix}
\dfrac{h_{2 \mathcal{T}}^{1} g_{\chi \mathcal{T}}+h_{2 u}^{1 2} g_{\chi u}^{2}}{\left(g_{\chi \mathcal{T}} \right)^{2}+\left(g_{\chi u}^{2} \right)^{2}}\dfrac{v_{2}}{v_{\chi}}	\\ \\
\dfrac{h_{1 \mathcal{T}}^{2} g_{\chi \mathcal{T}}+h_{1 u}^{2 2} g_{\chi u}^{2}}{\left(g_{\chi \mathcal{T}} \right)^{2}+\left(g_{\chi u}^{2} \right)^{2}}\dfrac{v_{1}}{v_{\chi}}	\\ \\
0
\end{pmatrix},
\end{equation}
and the SM matrix has the form:

\begin{equation}
\label{eq:SM-generic-mixing-matrix}
{V}^{U}_{\mathrm{SM}}=R_{23}(\theta_{23}^{U})
R_{13}(\theta_{13}^{U})R_{12}(\theta_{12}^{U}),
\end{equation}
with

\begin{subequations}
\label{subrotations-up}
\begin{align}
R_{12}(\theta_{12}^{U}) &= \begin{pmatrix}
c_{12}^{U}&	s_{12}^{U}	&	0\\	-s_{12}^{U}	&c_{12}^{U}	&0	\\	0&0&1
\end{pmatrix},	\\
R_{13}(\theta_{13}^{U}) &= \begin{pmatrix}
c_{13}^{U}	&0	&s_{13}^{U}	\\0	&1	&0	\\-s_{13}^{U}	&0	&c_{13}^{U}
\end{pmatrix},	\\
R_{23}(\theta_{23}^{U}) &= \begin{pmatrix}
1&0&0\\	0&	c_{23}^{U}	&s_{23}^{U}	\\0	&-s_{23}^{U}	&c_{23}^{U}
\end{pmatrix},
\end{align}
\end{subequations}
and $c_{ij}^{U}=\cos \theta_{ij}^{U}$ and $s_{ij}^{U}=\sin \theta_{ij}^{U}$. The angles $\theta_{ij}^{U}$ are specified by their tangents $t_{ij}^{U}=\tan \theta_{ij}^{U}$, which are \cite{1s3d-nonuniversal}:

\begin{equation}
\begin{split}
t_{12}^{U} &= \frac{h_{2 u}^{1 2} g_{\chi \mathcal{T}}-h_{2 \mathcal{T}}^{1} g_{\chi u}^{2}}{h_{1 u}^{2 2} g_{\chi \mathcal{T}}-h_{1 \mathcal{T}}^{2}g_{\chi u}^{2}} \frac{v_{2}}{v_{1}},\\
t_{13}^{U} &= \frac{h_{3 u}^{1 3} h_{1 u}^{3 3}+h_{3 u}^{1 1} h_{1 u}^{3 1}}{\left(h_{1 u}^{3 3}\right)^{2}+\left(h_{1 u}^{3 1}\right)^{2}}\frac{v_{3}}{v_{1}},\\
t_{23}^{U} &= 0.
\end{split}
\end{equation}

Finally, the squared mass eigenvalues are:

\begin{equation}
\label{eq:Up-Quarks-masses}
\begin{split}
m_{u}^{2}&=\frac{\left(h_{3 u}^{1 1}h_{1 u}^{3 3}-h_{3 u}^{1 3}h_{1 u}^{3 1}\right)^{2}}{(h_{1 u}^{3 3})^{2}+(h_{1 u}^{3 1})^{2}}
\frac{v_{3}^{2}}{2},	\\
m_{c}^{2}&=\frac{\left(h_{1 u}^{2 2}g_{\chi \mathcal{T}}-h_{1 \mathcal{T}}^{2}g_{\chi u}^{2}\right)^{2}}
{(g_{\chi \mathcal{T}})^{2}+(g_{\chi u}^{2})^{2}}\frac{v_{1}^{2}}{2},	\\
m_{t}^{2}&=\left[(h_{1 u}^{3 3})^{2}+(h_{1 u}^{3 1})^{2} \right]\frac{v_{1}^{2}}{2},\\
m_{T}^{2}& = \left[(g_{\chi \mathcal{T}})^{2}+(g_{\chi u}^{2})^{2} \right]\frac{v_{\chi}^{2}}{2}.
\end{split}
\end{equation}

\subsection*{Down sector}

The mass matrix of the down-type quarks is:

\begin{equation}
\label{eq:Down-mass-matrix}
\mathbb{M}_{D} = \frac{1}{\sqrt{2}}
\begin{pmatrix}
\Sigma_{d}^{11}	&	\Sigma_{d}^{12}	&	\Sigma_{d}^{13}	&	h_{1 \mathcal{J}}^{1 1}v_{1}	&	h_{1 \mathcal{J}}^{1 2}v_{1}	\\
h_{3 d}^{2 1}v_{3}	&	h_{3 d}^{2 2}v_{3}	&	h_{3 d}^{2 3}v_{3}	&	h_{2 \mathcal{J}}^{2 1}v_{2}	&	h_{2 \mathcal{J}}^{2 2}v_{2}	\\
h_{2 d}^{3 1}v_{2}	&	h_{2 d}^{3 2}v_{2}	&	h_{2 d}^{3 3}v_{2}	&	h_{3 \mathcal{J}}^{3 1}v_{3}	&	h_{3 \mathcal{J}}^{3 2}v_{3}	\\
0	&	0	&	0	&	g_{\chi \mathcal{J}}^{1}v_{\chi}	&0\\
0	&	0	&	0	&	0	&	g_{\chi \mathcal{J}}^{2}v_{\chi}
\end{pmatrix},
\end{equation}
where $\Sigma_{d}^{1k}$ are one-loop mass components. The see-saw matrix is:

\begin{equation}
\Theta^{D\dagger}_{L} = 
\begin{pmatrix}
\dfrac{h_{1 \mathcal{J}}^{1 1}}{g_{\chi \mathcal{J}}^{1}}\dfrac{v_{1}}{v_{\chi}}	&	
\dfrac{h_{1 \mathcal{J}}^{1 2}}{g_{\chi \mathcal{J}}^{2}}\dfrac{v_{1}}{v_{\chi}}	\\ \\
\dfrac{h_{2 \mathcal{J}}^{2 1}}{g_{\chi \mathcal{J}}^{1}}\dfrac{v_{2}}{v_{\chi}}	&	
\dfrac{h_{2 \mathcal{J}}^{2 2}}{g_{\chi \mathcal{J}}^{2}}\dfrac{v_{2}}{v_{\chi}}	\\ \\
\dfrac{h_{3 \mathcal{J}}^{3 1}}{g_{\chi \mathcal{J}}^{1}}\dfrac{v_{3}}{v_{\chi}}	&	
\dfrac{h_{3 \mathcal{J}}^{3 2}}{g_{\chi \mathcal{J}}^{2}}\dfrac{v_{3}}{v_{\chi}}
\end{pmatrix},
\label{seesaw-down}
\end{equation}
and the SM angles of $\mathbb{V}^{D}_{L,\mathrm{B}}$ are given by

\begin{equation}
\begin{split}
t_{12}^{D} &= \frac{{\Sigma}_{d}^{11}h_{3 d}^{2 1}+\Sigma_{d}^{12}h_{3 d}^{2 2}+\Sigma_{d}^{13}h_{3 d}^{2 3}}
{(h_{3 d}^{2 1})^2+(h_{3 d}^{2 2})^2+(h_{3 d}^{2 3})^2}\frac{1}{v_{3}},	\\
t_{13}^{D} &= \frac{\Sigma_{d}^{11}h_{2 d}^{3 1}+\Sigma_{d}^{12}h_{2 d}^{3 2}+\Sigma_{d}^{13}h_{2 d}^{3 3}}
{(h_{2 d}^{3 1})^2+(h_{2 d}^{3 2})^2+(h_{2 d}^{3 3})^2}\frac{1}{v_{2}},	\\
t_{23}^{D} &= \frac{h_{3 d}^{2 1}h_{2 d}^{3 1}+h_{3 d}^{2 2}h_{2 d}^{3 2}+h_{3 d}^{2 3}h_{2 d}^{3 3}}
{(h_{2 d}^{3 1})^2+(h_{2 d}^{3 2})^2+(h_{2 d}^{3 3})^2}\frac{v_{3}}{v_{2}},
\end{split}
\label{mixingangle-down}
\end{equation}
while the mass eigenvalues are:

\begin{equation}
\label{eq:Down-Quarks-masses-Light}
\begin{split}
m_{d}^{2} &= \frac{\left[
\left(\Sigma_{d}^{11}h_{3 d}^{2 2}-\Sigma_{d}^{12}h_{3 d}^{2 1} \right)h_{2 d}^{3 3} + 
\left(\Sigma_{d}^{13}h_{3 d}^{2 1}-\Sigma_{d}^{11}h_{3 d}^{2 3} \right)h_{2 d}^{3 2} + 
\left(\Sigma_{d}^{12}h_{3 d}^{2 3}-\Sigma_{d}^{13}h_{3 d}^{2 2} \right)h_{2 d}^{3 1}
\right]^2}{
\left[(h_{3 d}^{2 1})^2+(h_{3 d}^{2 2})^2 \right](h_{2 d}^{3 3})^2 + 
\left[(h_{3 d}^{2 3})^2+(h_{3 d}^{2 1})^2 \right](h_{2 d}^{3 2})^2 + 
\left[(h_{3 d}^{2 2})^2+(h_{3 d}^{2 3})^2 \right](h_{2 d}^{3 1})^2},	\\
m_{s}^{2} &= \frac{
\left[(h_{3 d}^{2 1})^2+(h_{3 d}^{2 2})^2 \right](h_{2 d}^{3 3})^2 + 
\left[(h_{3 d}^{2 3})^2+(h_{3 d}^{2 1})^2 \right](h_{2 d}^{3 2})^2 + 
\left[(h_{3 d}^{2 2})^2+(h_{3 d}^{2 3})^2 \right](h_{2 d}^{3 1})^2}
{(h_{2 d}^{3 3})^2+(h_{2 d}^{3 2})^2+(h_{2 d}^{3 1})^2}\frac{v_{3}^{2}}{2}, \\
m_{b}^{2} &= \left[(h_{2 d}^{3 3})^2+(h_{2 d}^{3 2})^2+(h_{2 d}^{3 1})^2 \right]\frac{v_{2}^{2}}{2},	\\
m_{J1}^{2} &= (g_{\chi \mathcal{J}}^{1})^2\frac{v_{\chi}^{2}}{2},	\quad 
m_{J2}^{2} = (g_{\chi \mathcal{J}}^{2})^2\frac{v_{\chi}^{2}}{2}.
\end{split}
\end{equation}

\subsection*{Charged lepton sector: left-handed}

The mass matrix of the charged leptons is:

\begin{equation}
\label{eq:Electron-mass-matrix}
\begin{split}
&\mathbb{M}_{E} =  \frac{1}{\sqrt{2}}
\begin{pmatrix}
0	&	h_{3 e}^{e \mu}v_{3}	&	0	&	h_{1 \mathcal{E}}^{e 1}v_{1}	&	0	\\
0	&	h_{3 e}^{\mu \mu}v_{3}	&	0	&	h_{1 \mathcal{E}}^{\mu 1}v_{1}	&	0	\\
h_{2 e}^{\tau e}v_{2}	&	0	&	h_{2 e}^{\tau \tau}v_{2}	&	0	&	0	\\
g_{\chi e}^{1 e}v_{\chi}	&	0	&	0	&	g_{\chi \mathcal{E}}^{1}v_{\chi}	&0\\
0	&	g_{\chi e}^{2 \mu}v_{\chi}	&	0	&0	&	g_{\chi \mathcal{E}}^{2}v_{\chi}
\end{pmatrix},
\end{split}
\end{equation}
with left-handed matrix rotations:

\begin{equation}
\Theta^{E\dagger}_{L} =\begin{pmatrix}
\dfrac{h_{1 \mathcal{E}}^{e 1}g_{\chi \mathcal{E}}^{1}v_{1}v_{\chi}}{2 m_{E^{2}}^{2}} &
\dfrac{h_{3 e}^{e \mu}g_{\chi e}^{2 \mu}v_{3}v_{\chi}}{2 m_{E^{1}}^{2}}\\
\dfrac{h_{1 \mathcal{E}}^{\mu 1}g_{\chi \mathcal{E}}^{1} v_{1}v_{\chi}}{2 m_{E^{2}}^{2}} &
\dfrac{h_{3 e}^{\mu \mu}g_{\chi e}^{2 \mu} v_{3}v_{\chi}}{2 m_{E^{2}}^{2}}	\\
\dfrac{h_{3 e}^{ e\mu}g_{\chi e}^{1 e} v_{3}v_{\chi}}{2 m_{E^{2}}^{2}} & 0
\end{pmatrix},
\label{see-saw-left-lepton}
\end{equation}
and

\begin{equation}
\label{eq:Electron-SM-Rotation-angles}
\begin{split}
t_{12}^{E_{L}} &\approx \frac{h_{1 \mathcal{E}}^{e 1}}{h_{1 \mathcal{E}}^{\mu 1}} ,\\
t_{13}^{E_{L}} &\approx \frac{g_{\chi \mathcal{E}}^{1}h_{3 e}^{e\mu }}
{g_{\chi e}^{ 1e}h_{1\mathcal{E}}^{ e 1} }\frac{v_{3}}{v_{1}},\\
t_{23}^{E_{L}} &\approx -\frac{2\left(g_{\chi \mathcal{E}}^{1}\right)^3 h_{3 e}^{e\mu }\left(h_{2 e}^{\tau \tau}\right)^{2}}
{\left(g_{\chi e}^{1e}\right)^3h_{1 \mathcal{E}}^{\mu 1} \left(h_{1 \mathcal{E}}^{e1}\right)^2 }\frac{v_{2}^2v_3}{v_1^{3}}.
\end{split}
\end{equation}
The mass values are:

\begin{equation}
\label{eq:Charged-Lepton-masses}
\begin{split}
m_{e}^{2} &= \frac{\left(h_{3 e}^{e \mu}h_{1 \mathcal{E}}^{\mu 1}-h_{3 e}^{\mu \mu}h_{1 \mathcal{E}}^{e 1}\right)^2}
{(h_{1 \mathcal{E}}^{e 1})^2+(h_{1 \mathcal{E}}^{\mu 1})^2}\frac{v_{3}^{2}}{2},		\\
m_{\mu}^{2} &= \frac{\left(h_{3 e}^{e \mu}h_{1 \mathcal{E}}^{e 1}+h_{3 e}^{\mu \mu}h_{1 \mathcal{E}}^{\mu 1}\right)^2}
{(h_{1 \mathcal{E}}^{e 1})^2+(h_{1 \mathcal{E}}^{\mu 1})^2}\frac{v_{3}^{2}}{2}+\frac{\left(h_{3e}^{ e\mu}\right)^2v_3^2}{2},	\\
m_{\tau}^{2} &= \left(h_{2 e}^{\tau \tau}\right)^{2}\frac{v_{2}^{2}}{2},	\\
m_{E{1}}^{2} &= \left[\left(g_{\chi \mathcal{E}}^{1}\right)^{2}+\left(g_{\chi e}^{1 e}\right)^{2}\right]\frac{v_{\chi}^{2}}{2},	\\
m_{E{2}}^{2} &= \left[\left(g_{\chi \mathcal{E}}^{2}\right)^2+\left(g_{\chi e}^{2\mu}\right)^2\right]\frac{v_{\chi}^{2}}{2}.
\end{split}
\end{equation}

\subsection*{Charged lepton sector: right-handed}

In addition, we need the rotations for the right-handed componente of the charged leptons. To obtain these parameters, we must construct the squared mass matrix $\mathbb{M}_R^{E}=\mathbb{M}_{E}^{\dag}\mathbb{M}_{E}$, which is diagonalized by the right-handed transformation $V_R^{E}$. In this case, the rotation matrix is expressed as:

\begin{eqnarray}
V_{R}^{E}=
\begin{pmatrix}
\Theta _{R11}^E & \Theta _{R12}^{ET}  \\
\Theta _{R21}^E  & \Theta _{R22}^E 
\end{pmatrix}
\begin{pmatrix}
V_{\text{SM}}^{E_{R}} & 0 \\
0 & V_{\text{new}}^{E_{R}}
\end{pmatrix},
\end{eqnarray}
with:

\begin{eqnarray}
\label{eq:righ-tlarge}
\Theta _{R11}^E&=&
\begin{pmatrix}
c_{14}^{E_{R}} & 0 & 0 \\
0 & c_{25}^{E_{R}} & 0\\
-s_{34}^{E_{R}}s_{14}^{E_{R}}  & 0 & c_{34}^{E_{R}}
\end{pmatrix} \nonumber \\
\Theta _{R12}^E&=&
\begin{pmatrix}
s_{14}^{E_{R}} & 0 & s_{34}^{E_{R}}c_{14}^{E_{R}} \\
0 & s_{25}^{E_{R}} & 0
\end{pmatrix} \nonumber \\
\Theta _{R21}^E&=&
\begin{pmatrix}
-c_{34}^{E_{R}}s_{14}^{E_{R}}  & 0 & -s_{34}^{E_{R}} \\
0 & -s_{25}^{E_{R}} & 0
\end{pmatrix} \nonumber \\
\Theta _{R22}^E&=&
\begin{pmatrix}
c_{34}^{E_{R}}c_{14}^{E_{R}} & 0 \\
0 & c_{25}
\end{pmatrix},
\end{eqnarray}
where the tangent of the mixing angles are:

\begin{eqnarray}
t_{25}^{E_{R}}&=&\frac{g_{\chi e}^{2\mu}}{g_{\chi \mathcal{E}}^{2}}, \nonumber \\
t_{34}^{E_{R}}&=&\frac{g_{\chi e}^{ 1e}}{g_{\chi \mathcal{E}}^{1}}, \nonumber \\
t_{14}^{E_{R}}&=&\frac{g_{\chi e}^{ 1e}}{\sqrt{\left(g_{\chi \mathcal{E}}^{1}\right)^2+\left(g_{\chi e}^{1e}\right)^2 } },
\label{right-miximg-angles-large}
\end{eqnarray}
while the SM mixing angles are:

\begin{eqnarray}
t_{12}^{E_{R}}&=&-\frac{g_{\chi e}^{1e}\left[\left(h_{1 \mathcal{E}}^{e1}\right)^2+\left(h_{1 \mathcal{E}}^{\mu1}\right)^2\right]}{g_{\chi \mathcal{E}}^{1}\left(h_{1 \mathcal{E}}^{e1}h_{3 e}^{e\mu}+h_{1 \mathcal{E}}^{\mu 1}h_{3 e}^{\mu \mu}\right)}\frac{v_1}{v_3}, \nonumber \\
t_{23}^{E_{R}}&=&\frac{g_{\chi \mathcal{E}}^{1}h_{2 e}^{\tau e}\left(h_{1 \mathcal{E}}^{e1}h_{3 e}^{ e\mu}+h_{1 \mathcal{E}}^{\mu1}h_{3 e}^{ \mu \mu}\right)}{g_{\chi e}^{1e} h_{2 e}^{\tau \tau}\left[\left(h_{1 \mathcal{E}}^{e1}\right)^2+\left(h_{1 \mathcal{E}}^{\mu 1}\right)^2\right]}\frac{v_3^2}{v_1v_2}, \nonumber \\
t_{13}^{E_{R}}&=&\frac{\left(g_{\chi \mathcal{E}}^{1}\right)^2h_{2 e}^{\tau e}h_{2 e}^{\tau \tau}}{\left(g_{\chi e}^{1e}\right)^2\left[\left(h_{1 \mathcal{E}}^{e1}\right)^2+\left(h_{1 \mathcal{E}}^{\mu 1}\right)^2\right]}\frac{v_2v_3}{v_1^2}.
\label{right-mixing-angles}
\end{eqnarray}

\subsection*{Natural parametrization}

In order to simplify the analysis, we separate the Yukawa interactions in three parts. First, the couplings among the ordinary SM fermions. Second, the interactions among the new particle content. Finally, the mixing couplings of the ordinary and the new particles. We assume a ''natural'' limit, where each part couple independently with the same strength. As a consequence, the mass matrices shares Yukawa couplings in some components. For example, in the up-type sector, by calling $h_{ku}^{ij}=h_u$, $g_{\chi \mathcal{T}}=g_{\mathcal{T}}$, $h_{i\mathcal{T}}^{j}=h_{\mathcal{T}}$ and $g_{\chi u}^2=g_u$, the mass matrix in (\ref{app:up-matrices}) become:

\begin{equation}
\mathbb{M}_{U} = \frac{1}{\sqrt{2}}
\begin{pmatrix}
h_{u}v_{3}	&	h_{u}v_{2}	&h_{u}v_{3}	&h_{\mathcal{T}}v_{2}	\\
0	&	h_{ u}v_{1}	&	0	&h_{ \mathcal{T}}v_{1}	\\
h_{ u}v_{1}	&	0	&	h_{ u}v_{1}	 &	0	\\
0	&	g_{ u}v_{\chi}	&	0	&	g_{ \mathcal{T}}v_{\chi}
\end{pmatrix}.
\label{app:up-matrices-2}
\end{equation}
In particular, in this limit, the mass of the top quark is:

\begin{eqnarray}
m_t^2=h_u^2v_1^2,
\label{top-mass}
\end{eqnarray}
from where we obtain the VEV of the first Higgs triplet, $v_1=m_t/h_u$. In the same form, the down-type mass matrix in (\ref{eq:Down-mass-matrix}) is written as

\begin{equation}
\label{eq:Down-mass-matrix-2}
\mathbb{M}_{D} = \frac{1}{\sqrt{2}}
\begin{pmatrix}
\Sigma_{d}^{11}	&	\Sigma_{d}^{12}	&	\Sigma_{d}^{13}	&	h_{ \mathcal{J}}v_{1}	&	h_{ \mathcal{J}} v_{1}	\\
h_{d}v_{3}	&	h_{d}v_{3}	&	h_{d}v_{3}	&	h_{ \mathcal{J}}v_{2}	&	h_{ \mathcal{J}}v_{2}	\\
h_{ d}v_{2}	&	h_{ d}v_{2}	&	h_{d}v_{2}	&	h_{ \mathcal{J}}v_{3}	&	h_{ \mathcal{J}}v_{3}	\\
0	&	0	&	0	&	g_{ \mathcal{J}}v_{\chi}	&0\\
0	&	0	&	0	&	0	&	g_{ \mathcal{J}}v_{\chi}
\end{pmatrix}.
\end{equation}
In this case, the masses of the quarks are:

\begin{eqnarray}
m_d^2&\sim &\Sigma _{d}, \nonumber \\
m_s^2&=&h_d^2v_3^2, \nonumber \\
m_b^2&=&\frac{3}{2}h_d^2v_2^2, \nonumber \\
m_{J}^2&=&\frac{1}{2}g_{ \mathcal{J}}^2v_{\chi}^2,
\end{eqnarray}
from where we obtain the VEVs for the other two Higgs triplets and the singlet as functions of the quarks masses: $v_2=\sqrt{2}m_b/\sqrt{3}h_d$, $v_3=m_s/h_d$ and $v_{\chi}=\sqrt{2}m_{J}/g_{ \mathcal{J}}$. With this scheme, the mixing angles (\ref{seesaw-down}) and (\ref{mixingangle-down}) can be parameterized as:

\begin{equation}
\Theta^{D\dagger}_{L} = \frac{h_{ \mathcal{J}}}{h_{ u}}
\begin{pmatrix}
\dfrac{1}{\sqrt{2}}\dfrac{m_t}{m_J}	&	
\dfrac{1}{\sqrt{2}}\dfrac{m_t}{m_J} \\ \\
\dfrac{1}{\sqrt{3}}\dfrac{m_b}{m_J}\dfrac{h_u}{h_d}	&	
\dfrac{1}{\sqrt{3}}\dfrac{m_b}{m_J}\dfrac{h_u}{h_d}		\\ \\
\dfrac{1}{\sqrt{3}}\dfrac{m_s}{m_J}\dfrac{h_u}{h_d}	&	
\dfrac{1}{\sqrt{3}}\dfrac{m_s}{m_J}\dfrac{h_u}{h_d}	
\end{pmatrix},
\label{seesaw-down-2}
\end{equation}
and

\begin{equation}
\begin{split}
t_{12}^{D} &= \frac{m_d}{m_s}	\\
t_{13}^{D} &= \frac{\sqrt{3}}{\sqrt{2}}\frac{m_d}{m_b}	\\
t_{23}^{D} &= \frac{\sqrt{3}}{\sqrt{2}}\frac{m_s}{m_b}.
\end{split}
\label{mixingangle-down-2}
\end{equation}
We see that the mixing matrix (\ref{seesaw-down-2}) depends on the ratio $r_{\mathcal{J}}=h_{ \mathcal{J}}/h_u$.

Regarding the lepton sector, the mass matrix (\ref{eq:Electron-mass-matrix}) simplify to:

\begin{equation}
\label{eq:Electron-mass-matrix-2}
\begin{split}
&\mathbb{M}_{E} =  \frac{1}{\sqrt{2}}
\begin{pmatrix}
0	&	h_{ e}v_{3}	&	0	&	h_{ \mathcal{E}}v_{1}	&	0	\\
0	&	h_{e}v_{3}	&	0	&	h_{ \mathcal{E}}v_{1}	&	0	\\
h_{ e}v_{2}	&	0	&	h_{ e}v_{2}	&	0	&	0	\\
g_{ e}v_{\chi}	&	0	&	0	&	g_{ \mathcal{E}}v_{\chi}	&0\\
0	&	g_{ e}v_{\chi}	&	0	&0	&	g_{ \mathcal{E}}v_{\chi}
\end{pmatrix},
\end{split}
\end{equation}
from where the charged lepton masses are expressed as:

\begin{eqnarray}
m_e^2&\approx &0, \nonumber \\
m_{\mu}^2 &=&\frac{3}{2}h_e^2v_3^2, \nonumber \\
m_{\tau}^2&=&\frac{1}{2}h_e^2v_2^2 \nonumber \\
m_{E}^2&=&\left[\left(g_e\right)^2+\left(g_{ \mathcal{E}}\right)^2\right]\frac{v_{\chi}^2}{2}.
\label{lepton-mass-natural}
\end{eqnarray}
Thus, the VEVs, in this case, can be written in terms of the lepton couplings as $v_2=\sqrt{2}m_{\tau }/h_e$, $v_3=\sqrt{2}m_{\mu }/\sqrt{3}h_e$ and $v_{\chi}=\sqrt{2}m_{E}/\sqrt{\left(g_e\right)^2+\left(g_{ \mathcal{E}}\right)^2}$.

For the mixing angles, we choose two of them as free parameters. For the left-handed angles in (\ref{eq:Electron-SM-Rotation-angles}), we choose $t_{13}^{E_L}$ as a free parameter, while for the right-handed angles in (\ref{right-miximg-angles-large}) we take $t_{25}^{E_R}$. Thus, with the natural parametrization, the other mixing angles are:

\begin{eqnarray}
t_{12}^{E_L}&\approx &1, \nonumber \\
t_{23}^{E_L}&\approx &-\frac{6m_{\tau}^2}{m_{\mu}^2}\left(t_{13}^{E_L}\right)^3, \nonumber \\
t_{12}^{E_R}&\approx &-\frac{1}{t_{13}^{E_L}}, \nonumber \\
t_{23}^{E_R}&\approx &\frac{m_{\mu }}{\sqrt{3}m_{\tau }}t_{13}^{E_L}, \nonumber \\
t_{13}^{E_R}&\approx &\frac{3m_{\tau }}{2m_{\mu }}\left(t_{13}^{E_L}\right)^2, \nonumber \\
t_{34}^{E_R}&\approx & t_{25}^{E_R}, \nonumber \\
t_{14}^{E_R}&\approx & s_{25}^{E_R},
\label{leptonmixing-natural}
\end{eqnarray} 
while the mixing matrix (\ref{see-saw-left-lepton}) takes the form:

\begin{equation}
\Theta^{E\dagger}_{L} =\frac{h_{ \mathcal{E}}}{h_u}\begin{pmatrix}
\dfrac{1}{\sqrt{2}}\dfrac{m_t}{m_{E}}c_{25}^{E_R} &
\dfrac{m_{\mu}}{m_{E}}\dfrac{h_{ \mathcal{E}}}{\sqrt{3}h_u}s_{25}^{E_R}  \\ \\
\dfrac{1}{\sqrt{2}}\dfrac{m_t}{m_{E}}c_{25}^{E_R}&
\dfrac{m_{\mu}}{m_{E}}\dfrac{h_{ \mathcal{E}}}{\sqrt{3}h_u}s_{25}^{E_R} \\ \\
\dfrac{m_{\mu}}{m_{E}}\dfrac{h_{ \mathcal{E}}}{\sqrt{3}h_u}s_{25}^{E_R} & 0
\end{pmatrix},
\label{see-saw-left-lepton-2}
\end{equation}
We also see that the above matrix is function of the ratio $r_{\mathcal{E}}=h_{ \mathcal{E}}/h_u$. 

Putting all the above matrices together, we will obtain each component of the original bi-unitary transformations $V^D_{L}$, $V^E_{L}$ and $V^E_{R}$. In particular, the neutral current couplings for electrons depends on $\left(V^E_{L}\right)_{31,41,51}$ and  $\left(V^E_{R}\right)_{21,41,51}$, while for muons we need $\left(V^E_{L}\right)_{32,42,52}$ and $\left(V^E_{R}\right)_{22,42,52}$. They are:

\begin{eqnarray}
\left(V^E_{L}\right)_{51,(52)}&=&0 \nonumber \\
\left(V^E_{L}\right)_{31,(32)}&=&\frac{-t_{13}^{E_L}}{\sqrt{2}\sqrt{1+36x^4\left(t_{13}^{E_L}\right)^6}}\left[c_{13}^{E_L}\pm 6x^2\left(t_{13}^{E_L}\right)^2\right] \nonumber \\
\left(V^E_{L}\right)_{41,(42)}&=&\frac{-1}{2}yr_{\mathcal{E}}c_{25}^{E_R}\left[s_{13}^{E_L}+ \frac{\mp1+6x^2c_{13}^{E_L}\left(t_{13}^{E_L}\right)^4}{\sqrt{1+36x^4\left(t_{13}^{E_L}\right)^6}} \right],
\label{lepton-left-biuni}
\end{eqnarray}
where $x=m_{\tau}/m_{\mu}$ and $y=m_t/m_E$, and:

\begin{eqnarray}
\left(V^E_{R}\right)_{21}&=&-t_{25}^{E_R}\left(V^E_{R}\right)_{51}=\frac{c_{13}^{E_L}c_{25}^{E_R}}{\sqrt{1+\frac{1}{3x^2}\left(t_{13}^{E_L}\right)^2}}\left[1-\frac{\sqrt{3}\left(t_{13}^{E_L}\right)^4}{2\sqrt{1+\frac{9}{4}x^2\left(t_{13}^{E_L}\right)^4}}\right] \nonumber \\
\left(V^E_{R}\right)_{22}&=&-t_{25}^{E_R}\left(V^E_{R}\right)_{52}=\frac{s_{13}^{E_L}c_{25}^{E_R}}{\sqrt{1+\frac{1}{3x^2}\left(t_{13}^{E_L}\right)^2}}\left[1+\frac{\sqrt{3}\left(t_{13}^{E_L}\right)^2}{2\sqrt{1+\frac{9}{4}x^2\left(t_{13}^{E_L}\right)^4}}\right], \nonumber \\
\left(V^E_{R}\right)_{41}&=&\frac{s_{13}^{E_L}s_{25}^{E_R}}{\sqrt{3}x\sqrt{1+\frac{1}{3x^2}\left(t_{13}^{E_L}\right)^2}}
\nonumber \\
&&+\frac{3s_{13}^{E_L}\left(t_{13}^{E_L}\right)^2s_{25}^{E_R}x}{2\sqrt{1+\frac{9}{4}x^2\left(t_{13}^{E_L}\right)^4}}\left[\frac{1}{\sqrt{1+\frac{1}{3x^2}\left(t_{13}^{E_L}\right)^2}}-\frac{c_{25}^{E_R}}{\sqrt{1+\left(s_{25}^{E_R}\right)^2}}\right] \nonumber \\
\left(V^E_{R}\right)_{42}&=&\frac{s_{13}^{E_L}s_{25}^{E_R}t_{13}^{E_L}}{\sqrt{3}x\sqrt{1+\frac{1}{3x^2}\left(t_{13}^{E_L}\right)^2}}
\nonumber \\
&&+\frac{3s_{13}^{E_L}t_{13}^{E_L}s_{25}^{E_R}x}{2\sqrt{1+\frac{9}{4}x^2\left(t_{13}^{E_L}\right)^4}}\left[\frac{-1}{\sqrt{1+\frac{1}{3x^2}\left(t_{13}^{E_L}\right)^2}}+\frac{c_{25}^{E_R}}{\sqrt{1+\left(s_{25}^{E_R}\right)^2}}\right]
\label{lepton-right-biuni}
\end{eqnarray}

\section*{Acknowledgment}
This work was supported by \textit{El Patrimonio Autonomo Fondo Nacional de Financiamiento para la Ciencia, la Tecnolog\'{i}a y la Innovaci\'on Francisco Jos\'e de Caldas} programme of COLCIENCIAS in Colombia. RM thanks to professor Germán Valencia for the kindly hospitality at Monash University and his useful comments.

\end{document}